\documentclass[12pt,a4paper]{amsart}

\usepackage{amssymb}
\usepackage{curves}

\makeatletter
\renewcommand{\@begintheorem}[2]{
\rm \trivlist \item [\hskip \labelsep {\bf #2\ \ #1.}]
                                }
\makeatother

\makeatletter

\DeclareFontFamily{U}{cyr}{}
\DeclareFontShape{U}{cyr}{m}{n}{
  <5> wncyr5 <6> wncyr6 <7> wncyr7 <8> wncyr8 <9> wncyr9 <10->
wncyr10}{}
\DeclareMathAlphabet{\mathcyr}{U}{cyr}{m}{n}

\input cyracc.def

\newcommand{\ZZ}{{\bf Z}}
\newcommand{\QQ}{{\bf Q}}
\newcommand{\RR}{{\bf R}}
\newcommand{\CC}{{\bf C}}

\newcommand{\NN}{{\bf N}} 
\newcommand{\FF}{{\bf F}}
\newcommand{\HH}{{\bf H}}
\newcommand{\PP}{{\bf P}}

\newcommand{\ccH}{{\mathcal H}}

\renewcommand{\gg}{{\mathfrak g}}
\newcommand{\hh}{{\mathfrak h}}
\renewcommand{\ss}{{\mathfrak s}}

\newcommand{\tomega}{{\tilde{\omega}}}
\newcommand{\bes}{\begin{equation*}}
\newcommand{\ees}{\end{equation*}}

\title{From qubits to E7}
\author{Bianca Letizia Cerchiai}
\author{Bert van Geemen}
\address{Dipartimento di Matematica, Universit\`a di Milano,
Via Saldini 50, I-20133 Milano, Italia}
\email{bianca.cerchiai@unimi.it}
\email{lambertus.vangeemen@unimi.it}

\begin{document}

\begin{abstract}
There is a intriguing relation between quantum information theory and super gravity, discovered by M.J.\ Duff and S.\ Ferrara. 
It relates entanglement measures for qubits to black hole entropy, 
which in a certain case involves the quartic invariant on the 
56-dimensional representation of the Lie group E7. In this paper we recall the relatively straightforward manner in which three-qubits lead to E7, or at least to the Weyl group of E7.  We also show how the Fano plane emerges in this context.
\end{abstract}

\maketitle

In this paper we consider some groups which come up naturally in the study of qubits and their state space. We show that the case of three-qubits naturally leads 
to the Weyl group of $E_7$. The methods used to show this link also provide a very natural interpretation for the emergence of the Fano plane 
when one restricts the $56$-dimensional representation of the complex 
Lie group $E_7(\CC)$ to seven (commuting) copies of $SL(2,\CC)$.

We view the $k$-qubits as the non-zero elements of the finite abelian group $L_k=(\ZZ/2\ZZ)^k$. The state space, denoted by $\ccH_k$,  is a $2^k$-dimensional complex vector space and there is a natural action of the qubits 
on this state space. This action can be extended to an action of the group generated by the generalized Pauli matrices. 
This group is the Heisenberg group $H_k$. 
It is non-abelian and has a quotient $V_k:=H_k/\CC^\times$ 
which is in a natural way a symplectic vector space of dimension $2k$ 
over the finite field $\FF_2=\ZZ/2\ZZ$.

The next actor to appear on stage is the normalizer $N_k$ of the Heisenberg group, viewed as a subgroup of $GL(\ccH_k)$. We will give explicit generators for this group.
The quotient of
$N_k$ by the subgroup $\CC^\times\cdot H_k$, consisting of scalar  multiples of the identity and the Heisenberg group, is the finite symplectic group 
$Sp(2k,\FF_2)$ (acting naturally on $V_k$). 

In the case $k=3$, there is a surjective homomorphism from the Weyl group 
$W(E_7)$ onto $Sp(6,\FF_2)$, with kernel just $\pm I$. We obtain this homomorphism in two ways, first by using Coxeter relations and second by
giving a surjective homomorphism $\pi:Q(E_7)\rightarrow V_3$ where
$Q(E_7)$ is the root lattice of $E_7$. This homomorphism is compatible with
the scalar product on $Q(E_7)$ and the symplectic form on $V_3$.
It maps the $63$ pairs of roots $\pm \alpha$ of $E_7$ to the $2^6-1=63$ non-zero elements of $V_3$. 
Moreover, the reflections $s_\alpha\in W(E_7)$ defined by the roots 
of $E_7$ correspond to the so-called transvections $t_v\in Sp(6,\FF_2)$. These transvections
are known to generate $Sp(6,\FF_2)$.

Having arrived at $W(E_7)$, we consider Lie subgroups isomorphic
to $SL(2,\CC)^7$ of $E_7(\CC)$. The choice of seven perpendicular roots of $E_7$
defines such a subgroup. Under the map $\pi$, this corresponds to giving $7$ points in $V_3\cong \FF_2^6$ which are the non-zero points in a Lagrangian subspace $L$ of $V_3$. Thus $L\cong \FF_2^3$ is a three dimensional vector space over $\FF_2$ and the associated projective space
is known as the Fano plane. At this point we felt we had to rederive, using the standard theory of highest weights, an important result from \cite{DFq}.
It relates the restriction to $SL(2,\CC)^7$  of the $56$-dimensional irreducible representation of $E_7(\CC)$ to the seven lines in the Fano plane.

To do so, we introduce the last actor: the Del Pezzo surface of degree two. 
These surfaces already made their appearance in String theory. Here we only use their Picard group in order to have a convenient notation for the roots and 
weights of $E_7$ involved in the computation.

In the appendices we discuss various other aspects, like symmetric and alternating forms on $\ccH_k$, the action of the normalizer $N_k$ on them and how this is related to quadratic forms on $V_k$. It is here that links with Hopf maps (see section \ref{hopf}),  Clifford algebras (see section \ref{clifford}) and the transformation formulas for theta functions (see also the Appendix of \cite{DG}) appear.

It is a pleasure to acknowledge stimulating discussions with Sergio Cacciatori and
Alessio Marrani. The latter in particular drew our attention to the paper
\cite{LSV} which also considers (finite) group theory in relation to qubits and Supergravity, see also our section \ref{lsv}. 

The first author's research is in part funded by the European grant PIRG-GA-2008-239412.

\section{From qubits to the Heisenberg group}

\subsection{Three-qubits}
In this section, three-qubits are simply considered as the $8$ elements of the group $L_3:=(\ZZ/2\ZZ)^3$.

\subsection{The three-qubit Hilbert space} \label{cH}
Starting from the group $L_3=(\ZZ/2\ZZ)^3$, we introduce the vector space $\ccH_3$ of $\CC$-valued maps $L_3\rightarrow\CC$. Notice that this vector space is isomorphic to $\CC^8$. A basis of $\ccH_3$ is given by the $8$ delta functions 
$$
\delta_x\,:\,L_3\,\longrightarrow\,\CC,\qquad \delta_x(y)=0\quad \mbox{if} \;x\neq y,
\quad \delta_x(x)=1,
$$
for $x,y\in L_3$.

\subsection{The action of $L_3$ on $\ccH_3$}
There is an obvious representation of $L_3$ on $\ccH_3$:
$$
U\,:\,L_3\,\longrightarrow\, \mbox{GL}(\ccH_3),\qquad (U_xf)(y):=f(x+y),
\qquad(x,y\in L_3,\; f\in\ccH_3).
$$
In the basis of the $\delta$-functions this action is  
$$
U_x\delta_y=\delta_{x+y}.
$$
Rather than write the corresponding $8\times 8$ matrices explicitly,
we consider a similar construction for $L_k:=(\ZZ/2\ZZ)^k$ where $k \in \NN$. 

For $k=1$, the vector space $\ccH_1$ has dimension $\sharp L_1=2$.
The matrices $U_0, U_1$ in the basis $\delta_0,\delta_1$ are, using the notation from \cite{LSV}, section 2, $(3)$:
$$
U_0\,=\,I\,=\,\left(\begin{array}{ll}1&0\\0&1\end{array}\right),\qquad
U_1\,=\,X\,:=\,\left( \begin{array}{cc}0&1\\1&0\end{array}\right).
$$
In the case $k=2$ one easily verifies that $\ccH_2=\CC^4=\CC^2\otimes\CC^2$,
with basis $\delta_{(a,b)}=\delta_a\otimes\delta_b$, and for $k=3$ one has similarly $\ccH_3=\CC^8=\CC^2\otimes\CC^2\otimes\CC^2$ and an element
$x=(a,b,c)\in L_3$ will act as $X^a\otimes X^b\otimes X^c$ with $a,b,c\in\ZZ/2\ZZ$, with the usual realization $\ZZ/2\ZZ=\{0,1\}$.

\subsection{A group action on $\ccH_k$}\label{grpact}
There is a natural way to extend this action to a bigger group: let $L_k^*$ be the dual group of $L_k$, that is, the group of homomorphisms 
$$
L_k^*\,:=\,\mbox{Hom}(L_k,\,\ZZ/2\ZZ)\quad(\cong L_k).
$$
Any element $x^*\in L_k^*$ is given by a linear form
$$
x^*(y):=x_1y_1+\ldots+x_ky_k \qquad (y=(y_1,\ldots,y_k)\in L_k)
$$ 
for a certain, uniquely determined, element $(x_1,\ldots,x_k)\in (\ZZ/2\ZZ)^k$,
we will simply write $x^*=(x_1,\ldots,x_k)$.

The group $L_k^*$ also acts naturally on $\ccH_k$. For convenience
we denote the action of $x^*\in L_k^*$ on $\ccH_k$ by $U_{x^*}$, so we abuse the definition of $U$ given earlier:
$$
U\,:\,L_k^*\,\longrightarrow\, \mbox{GL}(\ccH_k),\qquad (U_{x^*}f)(y):=(-1)^{x^*(y)}f(y),
\qquad(x^*\in L^*_k,\;y\in L_k,\;f\in\ccH_k).
$$
On the basis of $\delta$-functions of $\ccH_k$ one has $U_{x^*}\delta_y=x^*(y)\delta_y$
and  one obtains the matrices:
$$
U_{1^*}\,=\,Z\,:=\,\left( \begin{array}{cc}1&0\\0&-1\end{array}\right),\qquad
U_{x^*}\,=\,Z^a \otimes Z^b \otimes Z^c\qquad\mbox{for}\quad
x=(a,b,c)\in L_3.
$$

Now it is trivial to check that $X$ and $Z$ do not commute,
in fact $XZ=-ZX$. Thus there is no action of $L_k\times L_k^*$ on $\ccH_k$.
Rather a slightly bigger group acts, the Heisenberg group $H_k$, 
which is just the subgroup of $GL(\ccH_k)$ generated by the $U_x$, $U_{x^*}$ with $x\in L_k$ and $x^*\in L_k^*$. This group has order $2^{k+k+1}$ and contains also $-I$.
As we will see later, it is convenient to include also scalar multiplication by a fourth root of unity, i.e.\ the $sI$ with $s\in \mu_4$
where $\mu_4$ is the subgroup of fourth roots of unity
$$
\mu_4\,:=\,\{\pm 1,\pm i\}\qquad\subset\,\CC^{\times}
$$ 
of the multiplicative group of non-zero complex numbers $\CC^\times$. 
We now abuse the notation once more by defining:
$$
U_{(s,x,x^*)}\,=\,s U_{x^*} \, U_x\,=\, s (-1)^{x^*(x)} U_x U_{x^*}
\qquad\in GL(\ccH_k),\quad (s\in \mu_4,x\in L_k,x^*\in L_k^*).
$$
Note that the $U_x$ and $U_{x^*}$ do not commute in general,
this follows directly from the fact that $XZ=-ZX$.

The matrices $X,Z$ and $Y:=ZX=-XZ$ appear also in 
 \cite{LSV}, where they are called Pauli matrices.

\subsection{The Heisenberg group $H_k$}\label{Hk}
The linear maps $U_{(s,x,x^*)}$ which we just introduced can be seen as the representation matrices of an abstract group, the Heisenberg group $H_k$.

The Heisenberg group is defined as the set
$$
H_k=\mu_4 \times L_k\times L^*_k
$$
and the group operation is given by:
$$
(s,x,x^*)(t,y,y^*)\,:=\,(st(-1)^{y^*(x)},x+y,x^*+y^*) 
\quad  \mbox{for } s,t \in  \mu_4, \: x,y \in  L_k, \: x^*,y^* \in  L^*_k.
$$
One easily verifies that
$$
(s,x,x^*)^{-1}=(s^{-1}(-1)^{x^*(x)},x,x^*)
$$
(note that $x+x=0$(!))
and thus the commutator of two elements in $H_k$ is given by:
$$
(s,x,x^*)(t,y,y^*)(s,x,x^*)^{-1}(t,y,y^*)^{-1}=((-1)^{x^*(y)-y^*(x)},0,0).
$$

\subsection{The Schr\"odinger representation}
There is a (faithful) representation of the finite group $H_k$ on the vector space
$\ccH_k$, called the Schr\"odinger representation, defined, by abusing notation again, as follows:
$$
U:\,H_k\,\longrightarrow\,GL(\ccH_k),\qquad 
h=(s,x,x^*)\,\longmapsto\,U_h:=U_{(s,x,x^*)}.
$$
That $U$ is indeed an injective homomorphism follows easily from the fact that
$U_xU_{x^*}=(-1)^{x^*(x)}U_{x^*}U_x$.
More intrinsically, the Schr\"odinger representation can be defined as:
$$
\Bigl(U_{(s,x,x^*)} f\Bigr)(z) \,:=\,s(-1)^{x^*(z)} f(x+z)\qquad(f\in \ccH_k).
$$
In the basis described by the $\delta$-functions, this yields:
$$
U_{(s,x,x^*)}\delta_a\,:=\,s(-1)^{x^*(x+a)}\delta_{x+a}.
$$
Thus we recover the linear maps on $\ccH_k$ introduced in
section \ref{grpact}.

\subsection{A symplectic structure on $V_k=L_k \times L^*_k$}
\label{sympl}
The abelian group $L_k\times L^*_k\cong (\ZZ/2\ZZ)^{2k}$ can be recovered as the quotient of the Heisenberg group $H_k$ by its center, which coincides with $\mu_4$:
$$
0\,\longrightarrow\,\mu_4\,\longrightarrow\,H_k\,\longrightarrow \, 
L_k\times L^*_k\,\longrightarrow\,0.
$$
Note that $L_k\times L^*_k$ is a vector space over the field of two elements $\FF_2=\ZZ/2\ZZ$. This 
$2k$-dimensional $\FF_2$-vector space will be denoted by $V_k$:
$$
V_k\,:=\,L_k\times L_k^*.
$$
There is a natural symplectic form $E$ on $V_k$,
by this we mean a bilinear form $E$ which satisfies $E(v,v)=0$ for all $v\in  V_k$ and
which is non-degenerate (so for any non-zero $v$ there is a $w$ such  that $E(v,w)\neq 0$).
It is defined by:
$$
E\,:\,(L_k\times L^*_k)\,\times\,(L_k\times L^*_k)\,\longrightarrow\,\FF_2,
\qquad
E((x,x^*),(y,y^*))\,=\,y^*(x)-x^*(y).
$$

This symplectic form is closely related to the group structure of the
Heisenberg group. In fact, if we write (by another abuse of notation)
$$
(s,v)\,:=\,(s,x,x^*),\quad (t,w)\,:=\,(t,y,y^*)
\qquad
\mbox{with}\quad v=(x,x^*),w=(y,y^*)\in V_k,\;s,t\in\mu_4,
$$
then the commutator in $H_k$ can be written as:
$$
(s,v)(t,w)(s,v)^{-1}(t,w)^{-1}\,=\,(-1^{E(v,w)},\,0).
$$
In the Schr\"odinger representation we thus obtain:
$$
U_{(s,v)}U_{(t,w)}\,=\,(-1)^{E(v,w)}U_{(t,w)}U_{(s,v)}.
$$

\section{The normalizer of the Heisenberg Group}

\subsection{}
We have already seen how the qubits naturally lead to the Heisenberg group 
$H_k$ and its Schr\"odinger representation  on the vector space 
$\ccH_k$.
In this section we study a much bigger group $N_k$, 
the normalizer of $H_k$, which acts on $\ccH_k$. 
We show that the normalizer maps to the finite symplectic group
$Sp(2k,\FF_2)$
and we discuss various examples of elements in $N_k$. 
In particular, we show that the CNOT operators 
are elements of $N_3$ in section \ref{cnot} and we briefly discuss the subgroups $SL(3,\FF_2)\cong PSL(2,\FF_7)$ and $G_2(2)$ of
$Sp(6,\FF_2)$.
In section \ref{codes} we recall that  $N_1$ is related to the theory of codes.
In the case $k=3$ we relate $N_3$ to the Weyl group of $E_7$ in section \ref{nore7}.

\subsection{The normalizer of $H_k$}
\label{weyl}
For convenience, we will now identify the abstract group $H_k$
with its image under $U$ in $GL(\ccH_k)$.
The normalizer $N_k$ in $GL(\ccH_k)$ of the subgroup $H_k$ of $GL(\ccH_k)$, is defined as:
$$
N_k\,:=\,\{\,M\in\, GL(\ccH_k)\,:\,MH_kM^{-1}=H_k\,\}.
$$
An element $M\in N_k$ thus has the property that it maps an element $U_h\in H_k$ to another element $U_{h'}\in H_k$, we denote this map also by $M$,
$$
M:\,H_k\,\longrightarrow\, H_k,\qquad
U_h\,\longmapsto\,U_{h'}\qquad\mbox{if}\quad
MU_hM^{-1}\,=\,U_{h'}.
$$
Obviously, this map is an isomorphism of groups, that is, it is an automorphism of $H_k$:
$$
M(U_{h_1}U_{h_2})M^{-1}\,=\,
\bigl(MU_{h_1}M^{-1}\bigr)
\bigl(MU_{h_2}M^{-1}\bigr),\qquad(h_1,h_2\in H_k,\;M\in N_k).
$$
Note that each element of the center $\mu_4$ is mapped to itself ($M(sI)M^{-1}=sI$). A final observation is that if $M\in N_k$ then also $tM\in N_k$, for any $t\in \CC^\times$, but $M$ and $tM$ 
obviously give the same automorphism of $H_k$. 

The action of $M$ can be written as (with $U_v=U_{(1,x,x^*)}$ and $v=(x,x^*)\in V_k$):
$$
MU_vM^{-1}=i^{f_M(v)}U_{\phi_M(v)},\qquad (v\in V_k)
$$
for some  $f_M(v)\in \ZZ/4\ZZ$ and some map $\phi_M:V_k\rightarrow V_k$. 
We will discuss $\phi_M$ in section \ref{norsym}, 
but $f_M$ will not be of further interest.
As $M$ is an automorphism, we have in particular:
$$
MU_v^{-1}M^{-1}\,=\,i^{-f_M(v)}U_{\phi_M(v)}^{-1}.
$$

It should be emphasized that in the definition of $H_k$ it is essential that the center of $H_k$ is taken to be $\mu_4$, else the group $N_k$ will be smaller. 
In fact, since $M\in N_k$ defines an automorphism of $H_k$, 
it permutes the elements of a given order. 
As $U_v$ has order two if $v=(x,x^*)\neq 0$ and $x^*(x)=0$, whereas it has order four if $x^*(x)=1$, there is no element $M\in N_k$ which maps the elements of one type to those of the other type. However,
there are elements $M\in N_k$ such that $MU_vM^{-1}=\pm i U_w$ where $U_v^2=I$ and $U_w^2=-I$.

\subsection{The normalizer $N_k$ and the finite symplectic group}\label{norsym}
As $M\in N_k$ is an automorphism of $H_k$, the induced map 
$\phi_M$ on $V_k=H_k/\mu_4$ is also an automorphism, that is, 
it is a linear map on this vector space.
The remarkable thing  is that $\phi_M\in Sp(2k,\FF_2)$, the group of matrices preserving the symplectic form $E$ on $V_k$ (cf.\ section \ref{sympl}). This follows easily from the fact that $M$ is an automorphism of $H_k$ which acts as the identity on $\mu_4$:
{\renewcommand{\arraystretch}{1.7}
$$
\begin{array}{rcl}
U_{v}U_{w}U_{v}^{-1}U_{w}^{-1}&=&(-1)^{E(v,w)}\\
&=&M\bigl((-1)^{E(v,w)}\bigr)M^{-1}\\
&=&MU_{v}U_{w}U_{v}^{-1}U_{w}^{-1}M^{-1}\\
&=& \bigl(MU_{v}M^{-1}\bigr)
\bigl(MU_{w}M^{-1}\bigr)
\bigl(MU_{v}^{-1}M^{-1}\bigr)
\bigl(MU_{w}^{-1}M^{-1}\bigr)\\
&=& U_{\phi_M(v)}U_{\phi_M(w)}
U_{\phi_M(v)}^{-1}U_{\phi_M(w)}^{-1}\\
&=&(-1)^{E(\phi_M(v),\phi_M(w))}.
\end{array}
$$
}

It is well-known that the homomorphism
$$
N_k\,\longrightarrow\,Sp(2k,\FF_2),\qquad M\,\longmapsto\, \phi_M
$$
is surjective, in fact this is true not only for the case of $L_k$ but for any locally compact abelian group, cf.\ \cite{W}. We will construct explicit lifts of particular generators of $Sp(2k,\FF_2)$ to $N_k$ in section \ref{lift} below. 

Any element $U_w\in H_k$ is an element of $N_k$ since $U_wU_vU_w^{-1}=(-1)^{E(v,w)}U_v$, for all $v\in V$.
In particular, if $M=U_w\in N_k$ then $\phi_M=I$. 
The homomorphism above fits in an exact sequence:
$$
0\,\longrightarrow\,\CC^\times\cdot H_k\,\longrightarrow\,N_k\,\longrightarrow\,
Sp(2k,\FF_2)\,\longrightarrow 0.
$$

\subsection{Example: the case $k=1$}\label{exaN1}
It is worth getting more familiar with the normalizer in the simplest case $k=1$.
Observe that $Sp(2,\FF_2)$ is isomorphic to $SL(2,\FF_2)$ and is generated by the two elements 
$$
S\,=\,\left( \begin{array}{cc}0&1\\1&0\end{array}\right)\qquad\mbox{and}\quad T\,=\,\left( \begin{array}{cc}1&1\\0&1\end{array}\right).
$$
As $S({}^1_0)=({}^0_1)$ and $S({}^0_1)=({}^1_0)$,
to find the matrix (unique up to scalar multiple) $M_S$, 
it suffices to find a $2\times 2$ matrix $M_S$ such that
$$
M_S U_1=U_{1^*} M_S, \qquad M_S U_{1^*}=U_1 M_S
$$
(in fact, $U_1=U_{(1,1,0)}$ and $U_{1^*}=U_{(1,0,1)}$).
Similarly, to find $M_T$ one can impose :
$$
M_T U_1=U_1 M_T, \quad M_T U_{1^*}=-i U_{(1,1,1^*)} M_T,
$$
note that $U_{1^*}^2=I$ and $U_{(1,1,1^*)}^2=-I$.
These are linear equations for the coefficients of $M_S,M_T$ and one easily finds:
$$
M_S=\frac{1-i}{2} \left( \begin{array}{cc}1&1\\1&-1\end{array}\right),\qquad
M_T=\frac{1-i}{2} \left( \begin{array}{cc}1&i\\i&1\end{array}\right).
$$
The matrices $M_S,M_T$ are normalized in such a way that $|det M|=1$
and that the matrices have coefficients in the field $\QQ(i)$, that is, each coefficient is of the form $a+bi$ with $a,b\in\QQ$.

\subsection{CNOT operators}\label{cnot}\label{lsv}
The CNOT-operators are linear maps on $\ccH_3$, an example is
(cf.\ \cite{LSV} (19)):
$$
C_{12}\,:=\,
\left( \begin{array}{cccc}I \ &0 \ &0 \ &0 \ \\0&I&0&0\\0&0&0&I\\0&0&I&0\end{array}\right)
\qquad\in GL(\ccH_3)
$$
where the entries are $2\times 2$ blocks.
It is straightforward to check that for all $v\in V_3$ we have
$C_{12}U_vC_{12}^{-1}\in H_3$, it suffices to check this for the $6$ basis elements of $V_k$
of course. In particular, $C_{12}\in N_3$.
While checking, one finds for these basis elements that 
$C_{12}U_vC_{12}^{-1}=i^{f_{12}(v)}U_{\phi_{12}(v)}$, where $\phi_{12}\in Sp(6,\FF_2)$ is given by:
$$
\phi_{12}\,:=\,\left( \begin{array}{cc} A&0\\0&{}^tA^{-1}\end{array}\right),
\qquad \mbox{with}\quad
A\,:=\left( \begin{array}{rrr} 1 &\ 0 &\ 0 \\0&1&0\\0&1&1 \end{array}\right)\qquad
(\in GL(3,\FF_2)).
$$

Using the matrix with blocks $A,0,0,{}^tA^{-1}$ in $Sp(6,\FF_2)$ as above, 
for general $A\in GL(3,\FF_2)$, we
get an injective homomorphism $GL(3,\FF_2)\rightarrow Sp(6,\FF_2)$.
Its image is the subgroup of $Sp(6,\FF_2)$ which preserves the two subspaces
$L_3:=L_3\times\{0\}$ and $L_3^*:=\{0\}\times L_3^*$ of $V_3$ and thus this group is of particular interest for the qubits. It is easy to check that $GL(3,\FF_2)$ has $168=7\cdot 6\cdot 4$ elements (for the first column we can take any non-zero element $v_1$ of $\FF_2^3$, for the second column
any element $v_2$ not on the line $<v_1>=\{0,v_1\}$ and for the last column
any element $v_3$ not in the subspace $<v_1,v_2>=\{0,v_1,v_2,v_1+v_2\}$).
It is well known that $GL(3,\FF_2)\cong PSL_2(\FF_7)$.
This latter group was considered in Section 5 of  the paper \cite{LSV}. Note that $GL(n,\FF_2)=SL(n,\FF_2)$.

The finite Chevalley group $G_2(2)$ of order $12096$ 
is a (maximal) subgroup of $Sp(6,\FF_2)$, of index $120$. 
In section 4 of  \cite{LSV} 
a geometric configuration of points and lines in $\PP V_3=V_3-\{0\}$, the hexagon, 
is described which is invariant under the action of this group. 
There is an isomorphism $Sp(2k,\FF_2)\cong O(2k+1,\FF_2)$ 
(cf.\ \cite{Carter}, section 1.6), the latter group has already made 
its appearance in the link between quantum information and supergravity.

We would like to emphasize that 
the normalizer $N_3$ of the Heisenberg group $H_3$
and its quotient, the full group $Sp(6,\FF_2)$,
are the `natural' symmetry groups of the qubits.  

\subsection{Linear codes and weight polynomials}\label{codes}
As a minor digression, we show that the group $N_1$ is
closely related to weight polynomials of codes.
First of all, we renormalize the element $M_S$ with a factor $1/\sqrt{2}$ instead of $(1-i)/2$, to obtain the matrix $M'_S$ which satisfies $(M'_S)^2=I$:
$$
M'=M'_S\,:=\,\frac{1}{\sqrt{2}} \left( \begin{array}{cc}1&1\\1&-1\end{array}\right),
\qquad 
M'':=M'_SM_TM'_S\,=\,
\frac{1-i}{2} \left( \begin{array}{cc}1&0\\0&-i\end{array}\right).
$$
Note that $(M'')^2=diag(1,-1)$ and, cf.\ \cite{El} p.\ 1385, the matrices
$M',(M'')^2$ generate the dihedral group $D_{16}$ of order 16.
Moreover, the group $G_1$ generated by $M',M''$ is a group of order
$8\cdot 4!=192$ which maps onto $S_4$ with kernel the subgroup of diagonal matrices $\lambda I$, where  $\lambda\in \mu_8$, the group of $8$-th roots of unity (\cite{El}, p.\ 1386).
Note that $G_1\subset N_1$, in fact $\CC^\times G_1=N_1$.

The group $G_1$ acts naturally on polynomials in two variables $X,Y$. 
The weight enumerator $W_{H_8}$ of the Hamming code $H_8$ is the polynomial given by:
{\renewcommand{\arraystretch}{1.7}
$$
\begin{array}{rcl}
2W_{H_8}&:=&2(X^8+14X^4Y^4+Y^8)\\
&=&(X^2+Y^2)^4+(2XY)^4+(X^2-Y^2)^4\\
&=& Q[{}^0_0]^4+Q[{}^1_0]^4+Q[{}^0_1]^4,
\end{array}
$$
}
with polynomials $Q[{}^\epsilon_{\epsilon'}]$ as in section \ref{tensors}.
This polynomial is a $G_1$-invariant. 
Thus the normalizer of the Heisenberg 
group is related to coding theory. 
A natural explanation for this is via theta functions, 
see \cite{El}, p.\ 1388 and \cite{DG}.

\section{Transvections}

\subsection{}
We consider again the surjective homomorphism 
$N_k\rightarrow Sp(2k,\FF_2)$, where $N_k$ is the normalizer 
 of the Heisenberg group.
We introduce the transvections $t_v\in Sp(2k,\FF_2)$, where $v\in V_k$.
Such transvections in the symplectic group are similar to reflections in orthogonal groups.
The symplectic group is generated by these transvections (\cite{Ja}, $\S$ 6.9).
For $v\in V_k$ we give an explicit formula for an element 
$M_v$ of  $N_k$ such  that $M_v\mapsto t_v$.

\subsection{Transvections}\label{transvections}
The transvection $t_v$, for  $v \in V_k$, is the linear map:
$$
t_v: V_k \,\longrightarrow \,  V_k,\qquad w  \,\longmapsto\,  w + E(w, v)v.
$$
It is easy to verify  that transvections are indeed symplectic
$$
E(t_v(w),t_v(z))\,=\,E(w,z), \qquad \mbox{for }\quad v,w,z \in V_k
$$
using the bilinearity of $E$ and the fact that $E(v,v)=0$.

\subsection{Lifting transvections}\label{lift}
Given $v\in V_k$, there is a remarkably simple formula for a matrix $M_v \in N_k$
such that $M_v$ induces the transvection $t_v$, that is, 
$$
\phi:\,N_k\longrightarrow\,Sp(2k,\FF_2),\qquad
M_v\;\longmapsto \; \phi_{M_v}\,=\,t_v,\qquad(v\in V_k).
$$

For $v\in V_k$ we define:
$$
M_v=\left\{
\begin{array}{llr}
\frac{1-i}{2}\left(I+i U_v\right)\qquad& 
\mbox{if}\quad U_v^2=I&\mbox{equiv.,}\quad x^*(x)=0,\\ \\
\frac{1-i}{2}\left(I+U_v\right) \qquad& 
\mbox{if}\quad U_v^2=-I&\mbox{equiv.,}\quad x^*(x)=1.
\end{array}
\right.
$$

\

\noindent
Now assume that $U_v^2=-I$, to verify that indeed $M_v U_w M_v^{-1}=i^{f_M(w)} U_{t_v(w)}$ we use that 
$$
M_v^{-1}=\mbox{$\frac{1+i}{2}$} \left(I-U_v\right)\qquad \mbox{and}
\quad
U_vU_w=(-1)^{E(v,w)}U_wU_v.
$$
Therefore we get:
$$
\begin{array}{l}
M_v U_w M_v^{-1}\,=\,\frac{1}{2} \left(U_w + U_vU_w-U_wU_v - U_v U_w U_v\right)= \\ \\
=\left\{
\begin{array}{rrrll}
\frac{1}{2} \left(U_w+0-U_v^2U_w\right)=&
U_w=&U_{t_v(w)} &\! \mbox{ if } E(v,w)=0,\\ \\
\frac{1}{2} \left(U_w+2U_vU_w+U_v^2U_w\right)=&\!
(-1)^{y^*(x))} U_{v+w}=&\!(-1)^{y^*(x)} U_{t_v(w)} &\! \mbox{ if } E(v,w)=1.
\end{array}
\right.
\end{array}
$$

\

\noindent
The case $U_v^2=I$ can be handled similarly.
It is easy to check that $M_v^2=i^lU_v$ for some integer $l$,  depending on $v$. Together with $\CC^\times$, 
the $M_v$ thus generate the normalizer $N_k$.

\subsection{Examples of lifts}
To clarify how this construction works at the practical level, 
we give some examples in the case $k=1$.
For $v=(1,0)$ we are in the case $x^*(x)=0$ and
$$
U_v=U_{(1,1,0)}=U_1=\left( \begin{array}{cc}0&1\\1&0\end{array}\right), \quad U^2_1=I,\qquad\mbox{hence}\quad
M_1=\frac{1-i}{2} \left( \begin{array}{cc}1&i\\i&1\end{array}\right).
$$
Thus $M_1=M_T$ (cf.\ section \ref{exaN1}), 
this is not really surprising as
$$
t_v(w)\,=\,t_{(1,0)}(x,y)\,=\,(x,y)+y(1,0)=(x+y,y)\,=\,T(x,y)
$$
that is, $t_v=T\in Sp(2,\FF_2)$.
Analogously for $v=(0,1)$:
$$
U_{(1,0,1)}=U_{1^*}=\left( \begin{array}{cc}1&0\\0&-1\end{array}\right), \quad U^2_{1^*}=I \qquad\mbox{hence}\quad
M_{1^*}=\left( \begin{array}{cc}1&0\\0&-i\end{array}\right).
$$
Thus $M_{1^*}=(M'')^{-1}$ with $M''\in N_1$ as in section \ref{codes}.

\subsection{The normalizer $N_1$ and symmetric groups $S_3,S_4$}
The group $N_1$ maps onto $Sp(2,\FF_2)=SL(2,\FF_2)$. As $\FF_2^2$ has three non-zero elements and any $A\in SL(2,\FF_2)$ permutes these, we get a surjective homomorphism
$$
SL(2,\FF_2)\,\longrightarrow\,S_3,
$$
(if we number the elements of $\FF_2^2-\{0\}$ by
${\bf 1}=(1,0),{\bf 2}=(0,1)$ and ${\bf 3}=(1,1)$ then, with the notation as in \ref{exaN1}, we get 
$$
S\,\longmapsto\,(12),\qquad
T\,\longmapsto\,(23)\qquad(\in S_3)
$$ 
and these elements generate $S_3$).
The kernel of this homomorphism is $\CC^\times H_1$. The group
$N_1/\CC^\times$ still maps onto $S_3$ with kernel $H_1/\mu_4\cong(\ZZ/2\ZZ)^2$. Using the methods from \cite{El}
it is not hard to show that $N_1/\CC^\times\cong S_4$, a symmetric group, and that the subgroup $H_1/\mu_4$ corresponds to 
$\{e,(12)(34),(13)(24),(14)(23)\}$.

\section{Qubits and $E_7$}\label{cube7}

\subsection{} In this section we recall that $Sp(6,\FF_2)$ is a quotient of the Weyl group $W(E_7)$. 
Thus we went all the way from three-qubits and the associated Heisenberg group $H_3$ with its normalizer $N_3$ and quotient $N_3/(\CC^\times\cdot H_3)\cong Sp(6,\FF_2)$ to $W(E_7)$. 

Since  we had to discuss the root system $E_7$ in the first part of this section, 
we decided to recall also the results of Duff and Ferrara on the restriction of
the $56$-dimensional representation $V(\omega_7)$ of the complex Lie group $E_7(\CC)$ to $SL(2,\CC)^7$. 
There is also a nice finite geometric description of this 
which unfortunately is a bit long, so we only give it in  Appendix \ref{D}. 

\subsection{The root system $E_7$}
As usual, we fix a real vector space of dimension $7$, denoted simply by $\RR^7$, with a positive definite inner product $(-,-)$.
The set of $126$ roots of $E_7$ is then a subset of $\RR^7$. 
An explicit description of these roots is given in section \ref{dp2}.
The Dynkin diagram for $E_7$, reproduced in the section \ref{nore7},
codes the scalar products of the simple roots 
$\alpha_1,\ldots,\alpha_7\in \RR^7$ in the usual way: $(\alpha_i,\alpha_j)$ is $2$ if $i=j$, $-1$ if $\alpha_i,\alpha_j$ are connected by an edge and $(\alpha_i,\alpha_j)=0$ otherwise.

\subsection{Coxeter relations for transvections}\label{coxeter}
The transvections $t_v$, with $v\in V_k$, satisfy the following, easy to verify, relations:
$$
t_v^2=1,\qquad t_vt_w=t_wt_v\quad\mbox{if}\quad E(v,w)=0,\qquad
t_vt_wt_v=t_{v+w} \quad\mbox{if}\quad E(v,w)=1
$$
for $w\in V$. Note that if $E(v,w)=1$ then $(t_vt_w)^3=(t_vt_wt_v)(t_wt_vt_w)=t_{v+w}^2=I$ whereas $(t_vt_w)^2=I$ if $E(v,w)=0$.

\subsection{The normalizer $N_3$ and Weyl group $W(E_7)$}\label{nore7}
Using the Coxeter relations, it is easy to establish a relation between $Sp(6,\FF_2)$ and $W(E_7)$. In the diagram below, the simple roots $\alpha_i$ are numbered as in \cite{Bb}, p.264. Above the simple root $\alpha_i$ we wrote an element $v_i\in V_3$, not at random, but, as the reader should verify, 
in such a way that
$$
(\alpha_i,\alpha_j)\,\equiv\,E(v_i,v_j)\;\mbox{mod}\;2\qquad(1\leq i,j\leq 7).
$$

This relation between the inner product on the roots system $E_7$ and the symplectic form on $V_3$ implies in particular that $E(v,w)=0$ for points $v,w$ in the diagram except when they are connected by an edge (or the dotted edge), in which case $E(v,w)=1$.
The Weyl group $W(E_7)$ is isomorphic to the group generated by
seven elements $s_i$, with Coxeter relations: $s_i^2=I$, $s_is_j=s_js_i$ if $(\alpha_i,\alpha_j)=0$ and $(s_is_j)^3=I$ if $(\alpha_i,\alpha_j)=-1$.
As the $t_{v_i}\in Sp(6,\FF_2)$ satisfy the same Coxeter relations,
there is a surjective homomorphism from $W(E_7)$ to $Sp(6,\FF_2)$.
Comparing orders of the groups, one finds that the kernel has two elements
and one can verify (see also below) that the kernel is $\{\pm I\}\subset W(E_7)$.
The homomorphism $W(E_7)\rightarrow Sp(6,\FF_2)$ can also be obtained by a natural map from the root lattice
of $E_7$ to $V_3$ which we will give in section \ref{QV} and Appendix \ref{D}.

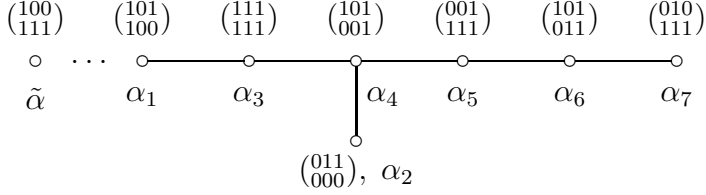
$$
 \begin{picture}(250, 95)%
   \put(-40,60){\circle{4}}%
   \put(-20,60){\makebox(0, 0)[c]{$\ldots$}}%
   \put(0, 60){\circle{4}}%
   \put(2, 60){\line(1, 0){36}}%
   \put(40, 60){\circle{4}}%
   \put(42, 60){\line(1, 0){36}}%
   \put(80, 60){\circle{4}}%
   \put(82, 60){\line(1, 0){36}}%
   \put(120, 60){\circle{4}}%
   \put(122, 60){\line(1, 0){36}}%
    \put(160, 60){\circle{4}}%
   \put(162, 60){\line(1, 0){36}}%
   \put(200, 60){\circle{4}}%
   \put(80, 58){\line(0, -1){26}}%
   \put(80, 30){\circle{4}}%
   \put(-40, 70){\makebox(0, 0)[b]{$({}^{100}_{111})$}}%
   \put(0, 70){\makebox(0, 0)[b]{$({}^{101}_{100})$}}%
   \put(40, 70){\makebox(0, 0)[b]{$({}^{111}_{111})$}}%
   \put(80, 70){\makebox(0, 0)[b]{$({}^{101}_{001})$}}%
   \put(120, 70){\makebox(0, 0)[b]{$({}^{001}_{111})$}}%
   \put(160, 70){\makebox(0, 0)[b]{$({}^{101}_{011})$}}%
   \put(200, 70){\makebox(0, 0)[b]{$({}^{010}_{111})$}}%
   \put(-40, 50){\makebox(0, 0)[t]{$\tilde{\alpha}$}}%
    \put(0, 50){\makebox(0, 0)[t]{$\alpha_1$}}%
    \put(40, 50){\makebox(0, 0)[t]{$\alpha_3$}}%
    \put(90, 50){\makebox(0, 0)[t]{$\alpha_4$}}%
    \put(120, 50){\makebox(0, 0)[t]{$\alpha_5$}}%
    \put(160, 50){\makebox(0, 0)[t]{$\alpha_6$}}%
   \put(200, 50){\makebox(0, 0)[t]{$\alpha_7$}}%
 \put(80, 25){\makebox(0, 0)[t]{$({}^{011}_{000}),\;\alpha_2$}}%
\end{picture}
$$

In the diagram above, $\tilde{\alpha}$ indicates the longest root (w.r.t.\ to the basis of simple roots) of $E_7$.
It is also the highest weight of the adjoint representation of $E_7$ and 
$\tilde{\alpha}$ is characterized by the scalar products
$$
(\tilde{\alpha},\alpha_1)=1,\quad (\tilde{\alpha},\alpha_i)=0,\quad(i=2,\ldots,7).
$$
One easily verifies that
$$
\tilde{\alpha}=2\alpha_1+2\alpha_2+3\alpha_3+4\alpha_4+3\alpha_5+2\alpha_6+\alpha_7.
$$

\subsection{The Weyl group $W(E_7)$}\label{we7}
The Weyl group of $E_7$ is the subgroup of $\RR^7$ generated by 
the reflections in the hyperplanes perpendicular to the roots of $E_7$. 
We will recall below that $-I$ is in $W(E_7)$.
The group $W(E_7)/<-I>$ is known to be simple and is isomorphic to $Sp(6,\FF_2)$, 
so one has an exact sequence:
$$
0\,\longrightarrow\,\mu_2\,\longrightarrow\,W(E_7)\,\longrightarrow\,
Sp(6,\FF_2)\,\longrightarrow\,0.
$$

{From} the diagram one can see that the orthogonal complement of the highest root 
$\tilde{\alpha}^\perp$ contains a root system of type $D_6$, which is
spanned by the six roots $\alpha_2,\alpha_3,\ldots,\alpha_7$.
This root system can be realized as the set of vectors $\pm(e_i\pm e_j)$ in $\RR^6$ with the standard inner product.
In particular, 
there are 6 mutually perpendicular vectors in $D_6$ (for example $e_1+e_2,e_1-e_2,e_3+e_4,\ldots,e_5-e_6$). 
Therefore there are $7$ perpendicular roots $\tilde{\alpha}=\beta_1,\ldots,\beta_7$ in $E_7$.

One immediate consequence of the existence of $7$ perpendicular roots in $E_7$ is the fact that $-I:\RR^7\rightarrow\RR^7$ is in the Weyl group of $E_7$. 
Indeed, the product of the seven reflections $s_{\beta_i}$ in $W(E_7)$ 
defined by the $7$ roots 
$\beta_1,\ldots,\beta_7$, which form an $\RR$-basis of $\RR^7$, 
is obviously $-I$.  

\subsection{The root lattice $Q(E_7)$ and $V_3$}\label{QV}
We now discuss another way to obtain the surjective homomorphism from
$W(E_7)$ to $Sp(6,\FF_2)$.

The root lattice of $E_7$ is denoted by $Q(E_7)\;(\cong\ZZ^7\subset\RR^7)$:
$$
Q(E_7)\,:=\,\{ n_1\alpha_1+\ldots+n_7\alpha_7\,\in\RR^7:\;n_i\,\in\,\ZZ\,\}.
$$
We define a group homomorphism
$$
\pi\,:Q(E_7)\,\longrightarrow\, V_3\cong \FF_2^6,\qquad
\pi(\alpha_i)\,:=\,v_i.
$$
Thus, for a general element in $Q(E_7)$ we have:
$$
\pi(n_1\alpha_1+\ldots+n_7\alpha_7)\,=\,n_1v_1+\ldots+n_7v_7.
$$
For example, since $\tilde{\alpha}=2\alpha_1+2\alpha_2+3\alpha_3+4\alpha_4+3\alpha_5+2\alpha_6+\alpha_7$
we get: 
$$
\pi(\tilde{\alpha})\,=\,\pi(\alpha_3)+\pi(\alpha_5)+\pi(\alpha_7)
\,=\,({}^{111}_{111})+({}^{001}_{111})+({}^{010}_{111})\,=\,
({}^{100}_{111}).
$$

As $(\alpha_i,\alpha_j)\equiv E(v_i,v_j)$ mod $2$ for
all $i,j$, the map $\pi$ is compatible with scalar product on the roots 
and the  symplectic form on $\FF_2^6$:
 $$
(\alpha,\beta)\,\equiv \, E(\pi(\alpha),\,\pi(\beta))\quad\mbox{mod}\,2\qquad
(\alpha,\beta\in Q(E_7)).
$$
The images of the $126$ roots of $E_7$ are exactly the
$63$ non-zero elements of $V_3\cong \FF_2^6$ (note that $\pi(\alpha)=\pi(-\alpha)$). 
The reflection defined by a root $\alpha$ is the linear map 
(we use $(\alpha,\alpha)=2$)
$$
s_\alpha:\,\RR^7\,\longrightarrow \,\RR^7,\qquad
x\,\longmapsto\,x-(x,\alpha)\alpha.
$$
If follows that this reflection induces the transvection $t_w$ in $w:=\pi(\alpha)$ 
because $t_w(v)=v+E(v,w)w$.
As $W(E_7)$ is generated by the reflections $s_\alpha$ and $Sp(6,\FF_2)$ is generated by the transvections, the map $\pi$ induces a surjective homomorphism $W(E_7)\rightarrow Sp(6,\FF_2)$. Comparing the orders of the groups, one finds that the kernel has two elements, and it is obvious that $-I$ is in the kernel, because it acts trivially on $V_3$. Thus $W(E_7)/\{\pm I\}\cong Sp(6,\FF_2)$. 

We refer to Appendix \ref{C} for a more intrinsic description of the map $\pi$.  

 \subsection{Copies of $SL(2,\CC)^7$ in $E_7(\CC)$.}\label{sl27}
 We illustrate that the map $\pi$ is quite useful in understanding Lie subgroups isomorphic to $SL(2,\CC)^7$ of the complex Lie group $E_7(\CC)$.

A positive root $\alpha$ of $E_7$ determines a subalgebra isomorphic to $sl(2)$ of the simple Lie algebra $\gg$ of type $E_7$. Usually its standard generators are denoted by $X_\alpha,
X_{-\alpha},H_\alpha$. We will write $sl(2)_\alpha$ for this subalgebra:
$$
sl(2)_\alpha\,=\,\langle\,X_\alpha,\,X_{-\alpha},\,H_\alpha\,\rangle\qquad
(\subset \gg\,=\,E_7).
$$
Recall that $X_{\pm\alpha}$ are a basis of the 1-dimensional
root spaces $\gg_{\pm\alpha}$, and that $H_\alpha:=[X_\alpha,X_{-\alpha}]$ lies in the Cartan algebra $\hh\cong \CC^7$ of $\gg$. 

Given two positive roots $\alpha,\beta$, the subalgebras 
$$
sl(2)_\alpha,\;sl(2)_\beta\quad\mbox{commute}
\quad\Longleftrightarrow\quad (\alpha,\beta)=0.
$$ 
In fact, as $(\alpha,\alpha)=2$ for all roots of $E_7$, 
we have $[H_\alpha,X_\beta]=(\alpha,\beta)X_\beta$, so this condition is necessary. Conversely, let $(\alpha,\beta)=0$. Then, as we just observed, $[H_\alpha,X_{\pm\beta}]=0=[H_\beta,X_{\pm\alpha}]$. 
Moreover, $\pm \alpha\pm \beta$ cannot be a root
since $(\pm\alpha\pm\beta,\pm\alpha\pm\beta)=4$,
whereas $(\alpha,\beta)\in\{\pm 2,\pm 1,0\}$ for any roots $\alpha,\beta$ of $E_7$. Therefore $\gg_{\pm\alpha\pm\beta}=0$ and as
$[X_{\pm\alpha},X_{\pm\beta}]\in \gg_{\pm\alpha\pm\beta}$, we conclude that the subalgebras commute. 

The root system of $E_7$ spans $\RR^7$, so there cannot be more than $7$ 
mutually orthogonal roots. We already observed that there are indeed sets of 
seven orthogonal roots in section \ref{we7}. 
Thus there are subalgebras isomorphic to $sl(2)^7$ in $\gg$ such that the direct sum of their Cartan algebras is the (chosen) Cartan algebra $\hh$ of $\gg$.
Equivalently, there is a (reducible) root subsystem of type $A_1^{\oplus 7}$ contained
in the root system $E_7$.

Using the map $\pi:Q(E_7)\rightarrow V_3$, which induces a bijection between the set of 63 pairs of roots $\pm \alpha$ and the non-zero points in $V_3$, 
it is easy to find all seven element subsets of perpendicular roots of $E_7$. 
In fact, since for roots $\alpha$ and $\beta\neq \pm\alpha$, we have 
$(\alpha,\beta)\in\{\pm 1,0\}$, the roots $\alpha,\beta$
are perpendicular, that is,
$$
(\alpha,\beta)\,=\,0\quad\Longleftrightarrow\, E(\pi(\alpha),\pi(\beta))\,=\,0
\qquad(\alpha,\beta\in E_7,\quad \beta\neq \pm\alpha).
$$
Note that given two such perpendicular roots, we immediately find a third root which is perpendicular to both, it is the root in $E_7$ which maps to 
$\pi(\alpha)+\pi(\beta)$.

A subspace $W$ of $V_3$ which is spanned by mutually orthogonal roots
is therefore an isotropic subspace: $E(v,w)=0$ for all $v,w\in W$.
As the symplectic form $E$ is non-degenerate, we must have $\dim W\leq 3$ and
$\dim W=3$ iff $\sharp(W-\{0\})=2^3-1=7$.
Thus we conclude that there is a bijection between the sets of $7$ perpendicular roots (up to sign) $\pm\beta_1,\ldots,\pm\beta_7$ and the
(non-zero points in a) Lagrangian (i.e.\ maximally isotropic) subspaces $W=\{0,\pi(\beta_1),\ldots,\pi(\beta_7)\}\cong\FF_2^3$ in $V_3$.

There are 135 Lagrangian subspaces in $V_3$ and thus there are $135$ such sets of roots, equivalently there are
$135$ root subsystems $A_1^{\oplus 7}\subset E_7$.
An elementary way to find this number is to notice that to find a basis of a Lagrangian subspace, one first chooses an arbitrary non-zero element $v_1\in V_3$, next an element $v_2\in v_1^\perp\subset V_3$, that is, $E(v_1,v_2)=0$, with $v_2\not\in <0,v_1>$ and finally a $v_3\in v_1^\perp\cap v_2^\perp$ with $v_3\not\in <0,v_1,v_2,v_1+v_2>$, this gives $(64-1)(32-2)(16-4)$ such bases. 
As $GL(3,\FF_2)$, which permutes the bases of a given subspace, has $168$ elements one finds that the number of Lagrangian subspaces is
$$
\frac{(64-1)(32-2)(16-4)}{168}\,=\,
\frac{(9\cdot 7 )(5\cdot 6)(4\cdot 3)}{7\cdot 6\cdot 4}
\,=\,9\cdot 5\cdot 3=135.
$$

\subsection{The Fano plane}\label{fano}
We drew a figure of such a Lagrangian subspace $L\cong\FF_2^3$, actually we only show the $7$ non-zero points. We also drew the codimension 1 subspaces as lines. 
Such a subspace consists of the points $\{0,a,b,a+b\}$,
and any two distinct points $a,b$ determine a unique line. 
Each such a line contains three non-zero points, note that one line is drawn as a circle. We also give the labels $A,\ldots,G$ for the points as in \cite{DFq}.

The knowledgeable reader will recognize this as the projective plane $\PP(L)$ over the field $\FF_2$, 
this projective plane is known as the Fano plane.
Like for the real and complex numbers, it is defined as
$$
\PP(L)\,:=\,(L-\{0\})/\FF_2^\times\;=\;L-\{0\},
$$
where we used that  $\FF_2=\{0,1\}$, so its multiplicative group $\FF_2^\times=\{1\}$ is trivial.  

\begin{figure}[ht]
\begin{center}
\setlength{\unitlength}{1cm}
\begin{picture}(8,9)(0,-1)
\put(0,0){\circle*{0.3}}
\put(8,0){\circle*{0.3}}
\put(4,6.93){\circle*{0.3}}
\put(4,0){\circle*{0.3}}
\put(4,2.3){\circle*{0.3}}
\put(2,3.47){\circle*{0.3}}
\put(6,3.47){\circle*{0.3}}
\put(0,0){\line(1,0){8}}
\put(0,0){\curve(0,0, 4,6.93)}
\put(0,0){\curve(4,6.93, 8,0)}
\put(4,0){\line(0,1){6.93}}
\put(0,0){\curve(2,3.47, 8,0)}
\put(0,0){\curve(6,3.47, 0,0)}
\put(4,2.3){\bigcircle{4.6}}
{\large \bf
\put(-1.7,-0.7){E=(0,0,1)}
\put(2.5,-0.7){F=(1,0,1)}
\put(6.5,-0.7){A=(1,0,0)}
\put(-0.8,3.36){B=(0,1,1)}
\put(6.3,3.36){G=(1,1,0)}
\put(3,2.2){D}
\put(4.4,2.2){(1,1,1)}
\put(2.5,7.3){C=(0,1,0)}
}
\end{picture}
\end{center}
\end{figure}
\setlength{\unitlength}{1pt}

\subsection{The restriction of $V(\omega_7)$ to 
$SL(2,\CC)^7$}\label{res56}
After briefly recalling some basic facts of representations of Lie groups and Lie algebras, we discuss
the restriction of the $E_7$-representation $V(\omega_7)$ to $SL(2)^7$
(cf.\ \cite{DFq}).

Recall that the weight lattice of $E_7$ is defined as
$$
P(E_7)\,:=\,
\{\,\omega\in \RR^7:\;(\omega,\alpha)\,\in\ZZ\;\forall\alpha\in Q(E_7)\,\}\,=\,
\{\omega\in \RR^7:\;(\omega,\alpha_i)\,\in\ZZ,\;i=1,\ldots,7\, \}.
$$
A weight $\omega$ is called dominant if $(\omega,\alpha_i)\geq 0$ for
all $i$.
There is a bijection $\omega\mapsto V(\omega)$ between the set of 
dominant weights of $E_7$ and the irreducible (finite dimensional) representations of 
the Lie algebra of $E_7$ (and of the complex Lie group $E_7(\CC)$). 

The (fundamental) dominant weight  $\omega_7$ is defined by the scalar products:
$$
(\omega_7,\alpha_7)=1,\quad (\omega_7,\alpha_i)=0,\quad(i=1,\ldots,6).
$$
One easily verifies that:
$$
\omega_7=(2\alpha_1+3\alpha_2+4\alpha_3+6\alpha_4+5\alpha_5+4\alpha_6+3\alpha_7)/2.
$$
Note that $\omega_7\not\in Q(E_7)$ due to the division by $2$. 
The irreducible representation $V(\omega_7)$ of the Lie algebra $\gg$ of type $E_7$ 
is known to have dimension $56$. 
It has a decomposition into weight spaces $V_\lambda$ with $\lambda\in P(E_7)$, 
so $Hv=\lambda(H)v$ for each $v\in V_\lambda$
where $H\in \hh$, the (chosen) Cartan algebra whose dual is $\RR^7\otimes\CC$.
Let $\Pi(\omega_7)$ be the set 
of $\lambda\in P(E_7)$ for which $V_\lambda\neq 0$.
It is known that  for each $\lambda\in \Pi(\omega_7)$ 
the weight space $V_\lambda$ is one dimensional and that the
$56$ weights $\lambda\in \Pi(\omega_7)$ are all in the $W(E_7)$-orbit of $\omega_7$. 
In particular, $\Pi(\omega_7)$ consists of $28$ pairs
of weights $\pm \lambda$.
$$
V(\omega_7)\,=\,\oplus_{\lambda\in \Pi(\omega_7)}\,V_\lambda,\qquad
\dim V_\lambda\,=\,1,\quad \sharp \Pi(\omega_7)=56.
$$

Let $\alpha$ be a positive root of $E_7$ and let $sl(2)_\alpha=\langle X_\alpha,X_{-\alpha},H_\alpha\rangle$ be the 
corresponding copy of $sl(2)$ as in section \ref{sl27}.
The action of $H_\alpha$ on the weight space
$V(\omega)_\lambda$ is (using again that $(\alpha,\alpha)=2$ 
for any root $\alpha$ of $E_7$):
$$
H_\alpha v\,=\,(\lambda,\alpha)v,\qquad \forall v\in V_\lambda.
$$
To find the integers $(\lambda,\alpha)$, that is the $sl(2)_\alpha$-weights, it is convenient to use the description of the roots and of the weights in $\Pi(\omega_7)$ given in sections \ref{rts},\ref{wts}.

The main result is that $(\lambda,\alpha)\in\{1,0,-1\}$, cf.\ section \ref{inps}.
Moreover 
for any positive root $\alpha$ there are exactly $32$ weights in $\Pi(\omega_7)$ 
with $(\lambda,\alpha)=0$ whereas there are $12$ weights 
with $(\lambda,\alpha)=1$ (and thus also $56-32-12=12$ weights with $(\lambda,\alpha)=-1$). 
The irreducible representations of $sl(2)$ are the $V(m)$, with $m\in \ZZ_{\geq 0}$ with
$\dim V(m)=m+1$. The weights of $V(m)$ are $m,m-2,\ldots,-m$. 

It follows that the only irreducible representations of 
$sl(2)_\alpha$ which occur in the restriction of $V(\omega_7)$ are $V(0)$, the 
trivial one-dimensional representation, and $V(1)$, the standard two dimensional 
representation. Moreover,
$$
V(\omega_7)_{|_{sl(2)_\alpha}}\,=\,V(0)^{n_0}\,\oplus\,V(1)^{n_1},\qquad
\left\{\begin{array}{rlr}
n_0&=\,\sharp\{\omega\in \Pi(\omega_7):\;(\omega,\alpha)=0\,\}\,=&32,\\
n_1&=\,\sharp\{\omega\in \Pi(\omega_7):\;(\omega,\alpha)=1\,\}\,=&12.
\end{array}
\right.
$$

This determines the restriction of $V\omega_7)$ to an $sl(2)_\alpha$. Now
let $L\subset V_3$ be the 
Lagrangian subspace corresponding to $SL(2,\CC)^7$.
To avoid confusion with the simple roots $\alpha_i$ we write
$$
L\,=\,\{0,\pi(\beta_1),\ldots,\pi(\beta_7)\,\},\qquad
(\beta_i\in E_7^+,\;(\beta_i,\beta_j)=0\;\mbox{if}\;i\neq j).
$$
The Lie algebra of $SL(2,\CC)^7$ will be denoted by
$$
\ss\,:=\,\oplus_{\beta\in L-\{0\}} \,sl(2)_\beta\,=\,\oplus_{i=1}^7sl(2)_{\beta_i}.
$$
The restriction of $V(\omega_7)$ must be a direct sum of tensor products
$$
V(\omega_7)_{|\ss}\,=\,\oplus_w \bigl( 
V(w_1)\boxtimes\ldots\boxtimes V(w_7)\bigr)^{n_w} 
\qquad(w=(w_1,\ldots,w_7)\in \{0,1\}^7).
$$
By explicit computation (see section \ref{res7}) or by using the finite geometry (see Appendix \ref{D}, \ref{res727}),
one obtains the following. Let $M\subset L$ be a codimension one subspace,
so $M\cong \FF_2^2$ and has three non-zero elements. We'll write
$$
M\,:=\,
\{0,\pi(\beta_i),\pi(\beta_j),\pi(\beta_k)=\pi(\beta_i)+\pi(\beta_j)\,\}
\quad\subset L. 
$$
Then, for each of the eight elements in $\{\pm 1\}^3$, there is exactly one weight
$\lambda\in \Pi(\omega_7)$ such that
$$
(\lambda,\beta_i)\,=\,\pm 1,\qquad (\lambda,\beta_j)=\pm 1,\qquad
(\lambda,\beta_k)=\pm 1,\qquad (\lambda,\beta_l)=0,\quad l\not\in\{i,j,k\}.
$$
This implies that 
there is a summand $V(w_1)\boxtimes\ldots\boxtimes V(w_7)$
which has $w_i=w_j=w_k=1$ and $w_l=0$ for $l\not\in\{i,j,k\}$.  
The dimension of this summand is $2^3\cdot 1^4=8$. 

As there are $7$ such lines $\PP(M)$ in the Fano plane and as
$\dim V(\omega_7)=56=8\cdot 7$, we conclude that
$$
V(\omega_7)_{|\ss}\,=\,\oplus_{\PP(M)\subset \PP(L)} V_M,\qquad
V_M:=V(w_1)\boxtimes\ldots\boxtimes V(w_7)\quad 
\mbox{with}\quad w_i=
\left\{\begin{array}{ll}
1\quad&\mbox{if}\;\pi(\beta_i)\in M,\\
0&\mbox{if}\;\pi(\beta_i)\not\in M,
\end{array}
\right.
$$
where the sum is over the seven lines $\PP(M)$ of the Fano plane $\PP(L)$.
This decomposition is written as
$$
56\,=\,(ABD)+(BCE)+(CDF)+(DEG)+(EFA)+(FGB)+(GAC)
$$
in \cite{DFq}, (5.1), where $A,B,\ldots, G$ denote the points and $(ABD)$,$\ldots$,$(GAC)$ are the seven lines
 in the Fano plane.

\section{Degree two Del Pezzo surfaces and $E_7$}

\subsection{}
We briefly recall the relation between the root system $E_7$ and the Del Pezzo surfaces of degree two. This leads to a convenient way to enumerate the roots of $E_7$ and the set of weights $\Pi(\omega_7)$
of the $56$-dimensional representation of $E_7(\CC)$.
In particular, we get a very explicit description of the restriction of $V(\omega_7)$ to $SL(2,\CC)^7$.
A reference for the Algebraic Geometry described here is \cite{DO}, in particular VII.4, p.120--123.

\subsection{The degree two Del Pezzo surfaces}\label{dp2}
Any degree two Del Pezzo surface $S$ can be realized as the blow up of the complex projective plane $\PP^2(\CC)$ in $7$ points $p_1,\ldots,p_7$ in general position. Its second cohomology group  is also its Picard group: $H^2(S,\ZZ)=Pic(S)\cong \ZZ^8$.
The intersection form on $H^2(S,\ZZ)$ is a bilinear form whose associated quadratic form has signature $(1+,7-)$, changing the sign we get a bilinear form on $Pic(S)$ which we will denote by $[-,-]$.
A basis of $Pic(S)$ is provided by the classes $e_0,e_1,\ldots, e_7$
where $e_0$ is the pull-back of the class of a line in $\PP^2$ and the $e_i$ are the classes of the exceptional divisors over the $p_i$, 
$$
Pic(S)\,=\,\ZZ e_0\,\oplus\,\ZZ e_1\,\oplus\,\ldots\,\oplus\,\ZZ e_7.
$$
This basis is orthogonal for the symmetric bilinear form $[-,-]$, one has:
$$
[e_0,e_0]\,=\,-1,\qquad [e_i,e_i]\,=\,1\quad(1\leq i\leq 7),\qquad
[e_i,e_j]\,=\,0\quad 0\leq i < j\leq 7.
$$
The canonical divisor class is given by
$$
K_S\,=\,-3e_0+e_1+\ldots+e_7,\qquad [K_S,K_S]\,=\,-2.
$$

\subsection{The roots of $E_7$}\label{rts}
The interesting thing for us is the orthogonal complement of $K_S$ in $Pic(S)$.
This is a lattice of rank $7$ which has a basis $d_1,\ldots,d_7$ such that the matrix $[d_i,d_j]$ has
determinant $-[K_S,K_S]=2$, in fact it is the root lattice $Q(E_7)$. Thus we find the roots of $E_7$ in an euclidean $\RR^7$ which lies in an $\RR^8\;(=Pic(S)\otimes\RR)$ with a Minkowski metric.

$$
 
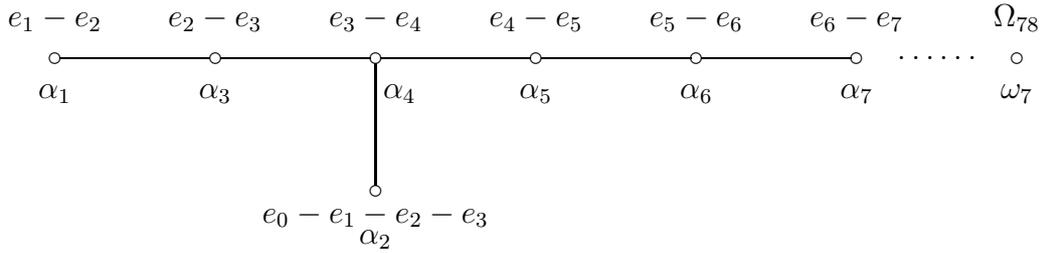
\begin{picture}(250, 100)%
  \put(270,80){\circle{4}}%
  \put(240,80){\makebox(0, 0)[c]{$\ldots\ldots$}}%
   \put(-90, 80){\circle{4}}%
   \put(-88, 80){\line(1, 0){56}}%
   \put(-30, 80){\circle{4}}%
   \put(-28, 80){\line(1, 0){56}}%
   \put(30, 80){\circle{4}}%
   \put(32, 80){\line(1, 0){56}}%
   \put(90, 80){\circle{4}}%
   \put(92, 80){\line(1, 0){56}}%
    \put(150, 80){\circle{4}}%
   \put(152, 80){\line(1, 0){56}}%
   \put(210, 80){\circle{4}}%
   \put(30, 78){\line(0, -1){46}}%
   \put(30, 30){\circle{4}}%
   \put(-90, 90){\makebox(0, 0)[b]{$e_1-e_2$}}%
   \put(-30, 90){\makebox(0, 0)[b]{$e_2-e_3$}}%
   \put(30, 90){\makebox(0, 0)[b]{$e_3-e_4$}}%
   \put(90, 90){\makebox(0, 0)[b]{$e_4-e_5$}}%
   \put(150, 90){\makebox(0, 0)[b]{$e_5-e_6$}}%
   \put(210, 90){\makebox(0, 0)[b]{$e_6-e_7$}}%
   \put(270, 90){\makebox(0, 0)[b]{$\Omega_{78}$}}%
\put(270, 70){\makebox(0, 0)[t]{$\omega_7$}}%
    \put(-90, 70){\makebox(0, 0)[t]{$\alpha_1$}}%
    \put(-30, 70){\makebox(0, 0)[t]{$\alpha_3$}}%
    \put(39, 70){\makebox(0, 0)[t]{$\alpha_4$}}%
    \put(90, 70){\makebox(0, 0)[t]{$\alpha_5$}}%
    \put(150, 70){\makebox(0, 0)[t]{$\alpha_6$}}%
   \put(210, 70){\makebox(0, 0)[t]{$\alpha_7$}}%
   \put(30, 25){\makebox(0, 0)[t]{$e_0-e_1-e_2-e_3$}}%
\put(30, 15){\makebox(0, 0)[t]{$\alpha_2$}}
\end{picture}
$$

It is easy to verify that the seven elements  $d_i$ written
on top of the $\alpha_i$ are a $\ZZ$-basis of $K_S^\perp$ and that they satisfy, for  $1\leq i,j\leq 7$:
$$
[d_i,d_j]\,=\,(\alpha_i,\alpha_j),\qquad 
d_1\,:=\,e_1-e_2,\quad d_2\,:=\,e_0-e_1-e_2-e_3,\quad\ldots,\quad
d_7\,=\,e_6-e_7.
$$
Thus $K_S^\perp\cong Q(E_7)$. The full list of the $63=21+35+7$ 
positive roots, with a convenient name,  is:
{\renewcommand{\arraystretch}{1.7}
$$
\begin{array}{rrcl}
R_{ij}=&e_i-e_j&\qquad& 1\leq i<j\leq 7,\\
R_{ijk8}=&e_0-e_i-e_j-e_k&&1\leq i<j<k\leq 7,\\
R_{i8}=&2e_0-(e_1+\ldots+e_7)+e_i&&1\leq i\leq 7.\\
\end{array}
$$
} 
With this notation, one has following property: if $I,J\subset\{1,\ldots,8\}$
are subsets with two or four elements, and $R_I,R_J$ denote the corresponding roots, then 
$$
[R_I,R_J]\,=\,\pm 1\qquad\Longleftrightarrow \quad \sharp \, (I\cap J)\,=\,1,3
$$
and $[R_I,R_J]\in \{0,2\}$ else.

\subsection{The weights in $\Pi(\Omega_7)$}\label{wts}
The $56$ weights in 
$\Pi(\omega_7)$ correspond to the $56$ exceptional curves of the
first kind on $S$ (these are smooth, rational curves $E$ on $S$ with
$E\cdot E=-1$ and $E\cdot K_S=-1$). The classes of these
$56=7+21+21+7$ curves in $Pic(S)$, with a name, are given by
{\renewcommand{\arraystretch}{1.7}
$$
\begin{array}{rrcl}
-\Omega_{i8}=&e_i&\qquad& 1\leq i\leq 7,\\
-\Omega_{ij}=&e_0-e_i-e_j&&1\leq i<j\leq 7,\\
\Omega_{ij}=&2e_0-(e_1+\ldots+e_7)+e_i+e_j&&1\leq i<j\leq 7,\\
\Omega_{i8}=&3e_0-(e_1+\ldots+e_7)-e_i&& 1\leq i\leq 7.
\end{array}
$$
} 
The first thing to notice is that $-\Omega_{ij}+\Omega_{ij}=K_S$,
and thus the classes $-\Omega_{ij},\Omega_{ij}$ do map to opposite elements in $K_S^\perp$.
One can verify that these are all the classes $l\in Pic(S)$ with $[l,K_S]=+1$ and $[l,l]=1$. Note that these classes do not lie in $K_S^\perp$, but we do have:
$$
[d_7,\Omega_{78}]\,=\,1,\quad[d_i,\Omega_{78}]\,=\,0\qquad(1\leq i\leq 6),\qquad\mbox{with}\quad
\Omega_{78}:=3e_0-(e_1+\ldots+e_6)-2e_7
$$
thus $\Omega_{78}$ projects onto $\omega_7\in P(E_7)$.

\subsection{Scalar products}\label{inps}
The scalar product between weights and roots is as follows:
for subsets $S,J\subset\{1,\ldots,8\}$, 
with $\sharp S=2$ and $\sharp J=2,4$, 
we have
$$
[\Omega_S,R_J]\,=\,\pm 1\qquad \Longleftrightarrow \quad 
\sharp\,(S\cup J)\,-\,\sharp(S\cap J)\,\equiv \,2\;\mbox{mod}\;4.
$$

It is now easy to verify that indeed (cf.\ section \ref{res56})
$$
[\Omega_S,R_J]\in \{1,0,-1\}\qquad\forall \Omega_S\in\Pi(\omega_7),\;R_J\in \Phi(E_7)^+.
$$ 
It is also easy to see that the weights 
$\lambda=\pm\Omega_S\in\Pi(\omega_7)$ with 
$[\pm\Omega_S,R_{12}]=0$ are:
$$
[\pm\Omega_S,R_{12}]\,=\,0\quad\Longrightarrow \;
S\,=\,\{i,j\},\;3\leq i<j\leq 7;\quad\{i,8\},\;3\leq i\leq 7;\quad \{12\};
$$
so we find $2(10+5+1)=32$ such  weights.

\subsection{The action of $W(E_7)$}
The action of a reflection $s_{\alpha}$ in the Weyl group is induced by the reflections $s_d$ on $Pic(S)$, where as usual $s_d(e):=e-[e,d]d$ for an element
$d\in Pic(S)$ with $[d,d]=2$.
This leads to the simple (and easily verified) formulas:
$$
s_{e_i-e_j}\,:\,\left\{
\begin{array}{lcrl}
e_0&\longmapsto& e_0,&\\ 
e_i  &\longmapsto& e_j,&\\
e_j  &\longmapsto& e_i,&\\
e_k  &\longmapsto& e_k&\mbox{if}\;k\neq i,j,
\end{array}
\right.
$$
thus these reflections induce permutations of the indices. The reflection in the simple root $\alpha_2$ is only slightly more complicated:
$$
s_{e_0-e_1-e_2-e_3}\,:\,\left\{
\begin{array}{lcrl}
e_0&\longmapsto& 2e_0-e_1-e_2-e_3,&\\ 
e_i  &\longmapsto& e_0-e_j-e_k&\mbox{if}\;k\in\{ 1,2,3\}\:\mbox{and}\;\{i,j,k\}=\{1,2,3\},\\
e_k  &\longmapsto& e_k&\mbox{if}\;k\not\in\{ 1,2,3\}.
\end{array}
\right.
$$

\subsection{The restriction of $V(\omega_7)$}\label{res7}
In the table below, we give both the labels $A,\ldots,G$ from \cite{DFq}, Figure 2 and from our figure of the Fano plane
in section \ref{fano}
(to be precise, we wrote $({}^{abc}_{000})$ for $(a,b,c)$). We also give the corresponding root $R_J$ of $E_7$,
that is, $\pi(R_J)=({}^{abc}_{000})$. 

For example, one writes $R_{2568}$ as a linear combination of $\alpha_1=e_1-e_2$, $\ldots$, $\alpha_7=e_6-e_7$: 
$$
e_0-e_2-e_5-e_6\,=\,(e_0-e_1-e_2-e_3)+(e_1-e_2)+(e_2-e_3)+2(e_3-e_4)+2(e_4-e_5)+(e_5-e_6)
$$
that is, in the notation of section \ref{rts}:
$$
R_{2568}\,=\,\alpha_2+\alpha_1+\alpha_3+2\alpha_4+2\alpha_5+\alpha_6.
$$
Using the map $\pi$ from section \ref{QV}, which maps $\alpha_i\mapsto v_i$, we then have:
$$
\pi(R_{2568})\,=\,({}^{011}_{000})+({}^{101}_{100})+({}^{111}_{111})+({}^{101}_{011})\,=\,
({}^{100}_{000}).
$$

{\renewcommand{\arraystretch}{1.5}
$$
\begin{array}{|@{\hspace{6pt}}c|@{\hspace{6pt}}c| @{\hspace{6pt}}c|@{\hspace{6pt}}c|@{\hspace{6pt}}c| @{\hspace{6pt}}c|@{\hspace{6pt}}c|} \hline
A&C&E&G&B&F&D\\ \hline
({}^{100}_{000}) &({}^{010}_{000})&
({}^{001}_{000})&({}^{110}_{000})&
({}^{011}_{000})&
({}^{101}_{000})&({}^{111}_{000})\\ \hline
R_{2568}&R_{3468}&R_{3578}&R_{1678}&
R_{1238}&R_{1458}&R_{2478}
\\ \hline
\end{array}
$$
} 

The next table gives, for each line in this Fano plane, 
the four weights $\Omega_S$ of the representation $V(\omega_7)$ 
such that $\pm \Omega_S$ has
inner product $\pm 1$ with the roots
corresponding to the three points on the line
and which are perpendicular to the other four roots.

In this table one should observe that the three roots $R_I,R_J,R_K$ which correspond to points on a line $\PP(M)$ 
in the Fano plane $\PP(L)$ have  index sets of the type 
$$
I\,=\,\{a,p,q,8\},\quad J\,=\,\{a,r,s,8\},\quad K\,=\,\{a,u,v,8\},\qquad\{a,p,q,r,s,u,v\}=\{1,\ldots,7\},
$$
so the common index $a$ labels the lines. The four pairs of weights $\pm \Omega_S$ which have scalar product $\pm 1$ with each of these three roots have  $S=\{p,q\},\{r,s\},\{u,v\},\{a,8\}$.
Indeed, one has $\sharp (S\cup I)-\sharp (S\cap I)=6-0=6$ or $4-2=2$ 
hence $[\Omega_S,R_I]=\pm 1$.

{\renewcommand{\arraystretch}{1.4}
$$
\begin{array}{|@{\hspace{6pt}}l||@{\hspace{4pt}}c |@{\hspace{6pt}}c @{\hspace{6pt}}   |@{\hspace{6pt}}c @{\hspace{6pt}}|}
\hline P(M)& a&
\mbox{roots in $M$}&
\mbox{weights}
\\\hline \hline
FGB&1&R_{1678},R_{1238},R_{1458}  & 
\Omega_{23},\Omega_{45},\Omega_{67},\Omega_{18}
\\ \hline
ABD&2&R_{2568},R_{1238},R_{2478}&
\Omega_{13},\Omega_{47},\Omega_{56},\Omega_{28}
\\ \hline
BCE&3&R_{3468},R_{3578},R_{1238}&
\Omega_{12},\Omega_{46},\Omega_{57},\Omega_{38}
\\ \hline
CDF&4&R_{3468},R_{2478},R_{1458} &
\Omega_{15},\Omega_{27},\Omega_{36},\Omega_{48}
\\ \hline
EFA&5&R_{2568},R_{3578},R_{1458}&
\Omega_{14},\Omega_{26},\Omega_{37},\Omega_{58}
\\ \hline
GAC&6&R_{2568},R_{3468},R_{1678}&
\Omega_{17},\Omega_{25},\Omega_{34},\Omega_{68}
\\ \hline
DEG&7&R_{3578},R_{1678},R_{2478}   & 
\Omega_{16},\Omega_{24},\Omega_{35},\Omega_{78}
\\ \hline
\end{array}
$$
} 

\section{Concluding remarks}
\subsection{The Weyl group}
The Weyl group of $E_7(\CC)$ can also be defined as the quotient
group
$$
W(E_7)\,=\,N(T)/T
$$
where $T\subset E_7(\CC)$ is a maximal torus and $N(T)$ is the normalizer of $T$:
$$
N(T)\,:=\,\{A\in E_7(\CC)\,:\,ATA^{-1}\,=\,T\,\},\qquad
T\cong (\CC^\times)^7\,\subset\,E_7(\CC).
$$
There seems to be a close relation between $N(T)$ and the image 
of $N_3$, the normalizer of the Heisenberg group, in 
$GL(V(\omega_7))\cong GL(\wedge^2\ccH\oplus\wedge^2\ccH^*)$.

\subsection{The Del Pezzo surfaces}
A rather deep link between Del Pezzo surfaces of degree two and the algebraic variety defined by quartic invariant on $V(\omega_7)$,
involving the Cox ring of the Del Pezzo surfaces, was recently established by Derenthal \cite{D}.

\subsection{Qutrits and $E_6$}
A variant of the qubits are the $k$-qutrits, which are elements of
$\FF_3^k$ (here $\FF_3$ is the field with three elements, $\FF_3\cong \ZZ/3\ZZ$).
In \cite{DF}, the restriction of the $27$-dimensional representation of $E_6(\CC)$ 
to a subgroup $SL_3(\CC)^3$ was considered in this context. 
An inclusion of three perpendicular copies of the root system $A_2$ into the root system $E_6$ determines such a subgroup. 

It may be of interest to know that there 
are $40$ such inclusions (similar to the $135$ choices of 
seven orthogonal copies of $A_1$ in $E_7$).
These subsystems actually have come up in the description of the 
moduli space of marked cubic surface (i.e.\ Del Pezzo surfaces of degree three).
This moduli space has 40 cusps (special boundary points, they can be blown up to give the so-called cusp divisors), which correspond to such root subsystems of $E_6$, see for example \cite{CG}. That paper (section 7.4) uses a well-known surjective map $Q(E_6)\rightarrow (\ZZ/3\ZZ)^5$ which might be useful for 
the study of the link between qutrits and $E_6$.

\

\appendix
\section{}
\subsection{} In this appendix we consider symmetric and antisymmetric bilinear forms on the state space $\ccH$. The action of the Heisenberg group singles out particular forms which are of interest for example for the Hopf maps.

\subsection{The action of $H_k$ on tensors}\label{tensors}
The Schr\"odinger representation $U:H_k\rightarrow GL(\ccH_k)$
defines representations 
$$
U^{<n>}\,:\,H_k\,\longrightarrow\, GL(\ccH_k^{\otimes n})
$$
on the $n$-fold tensor product of $\ccH_k$ with itself.
As the commutator of two elements in $H_k$ is $\pm 1$,
the action of $H_k$ on $\ccH_k^{\otimes n}$ factors over the abelian group $(\ZZ/2\ZZ)^{2k}$ in the case $n$ is even.

In the case $n=2$, this representation splits into the representations on the symmetric and anti-symmetric tensors, which will be of particular interest to us:
$$
\ccH_k^{\otimes 2}\,=\,S^2(\ccH_k)\,\oplus\,\wedge^2(\ccH_k),\qquad
\dim S^2(\ccH_k)=2^{k-1}(2^k+1),\quad 
\dim \wedge^2(\ccH_k)=2^{k-1}(2^k-1).
$$

This case is particularly nice, as $\dim \ccH_k\otimes\ccH_k=2^{2k}$
and each one dimensional irreducible representation of $(\ZZ/2\ZZ)^{2k}$ occurs once in this space. 
We now give the explicit decomposition of $S^2(\ccH_k)$ and $\wedge^2(\ccH_k)$ into irreducible subrepresentations.

The symmetric tensors can be written more conveniently as polynomials in variables $X_\sigma$, $\sigma \in L_k$, where
$X_\sigma=\delta_\sigma$. The elements of 
$S^n(\ccH_k)$ are then homogeneous polynomials of degree $n$ in these variables. We have:
$$
S^2(\ccH_k)\,=\,\oplus_{(\epsilon,\epsilon'),\;\epsilon'(\epsilon)=0}\;
\CC Q[{}^\epsilon_{\epsilon'}],
$$
with
$$
Q[{}^\epsilon_{\epsilon'}]\,:=
\sum_{\sigma \in L_k} (-1)^{\epsilon'(\sigma)} X_\sigma X_{\sigma+\epsilon},
$$
where $(\epsilon,\epsilon')\in L_k\times L_k^*$
and the $X_\sigma=\delta_\sigma$ are the delta functions.

Here are some concrete examples of these polynomials:
in the case $k=1$ one has the three polynomials:
$$
Q[{}^0_0]\,:=\,X_0^2\,+\,X_1^2,\qquad
Q[{}^0_1]\,:=\,X_0^2\,-\,X_1^2,\qquad
Q[{}^1_0]\,:=\,2 X_0 X_1,
$$
while for $k=2$ one has the $10$ polynomials:
$$
\begin{array}{ll}
Q[{}^{00}_{00}]\,:=\,X_{00}^2\,+\,X_{01}^2+\,X_{10}^2+\,X_{11}^2,&
Q[{}^{01}_{00}]\,:=\,2 \left( X_{00} X_{01}\,+\,X_{10} X_{11} \right),\\
Q[{}^{00}_{11}]\,:=\,X_{00}^2\,-\,X_{01}^2-\,X_{10}^2+\,X_{11}^2,&
Q[{}^{11}_{00}]\,:=\,2\left( X_{00} X_{11}\,+\,X_{01} X_{10}\right)\\
Q[{}^{00}_{10}]\,:=\,X_{00}^2\,+\,X_{01}^2-\,X_{10}^2-\,X_{11}^2,&
Q[{}^{01}_{10}]\,:=\,2\left( X_{00} X_{01}\,-\,X_{11} X_{10}\right),\\
Q[{}^{00}_{01}]\,:=\,X_{00}^2\,-\,X_{01}^2+\,X_{10}^2-\,X_{11}^2&
Q[{}^{10}_{01}]\,:=\,2\left( X_{00} X_{10}\,-\,X_{11} X_{01}\right),\\
Q[{}^{10}_{00}]\,:=\,2\left( X_{00} X_{10}\,+\,X_{11} X_{01}\right),&
Q[{}^{11}_{11}]\,:=\,2\left( X_{00} X_{11}\,-\,X_{10} X_{01}\right).
\end{array}
$$

To verify that we do have a decomposition of $S^2(\ccH_k)$ into
$H_k$-representations, the crucial observation is that 
these spaces are 
invariant under the action of $H_k$:
$$
U_{(s,x,x^*)}Q[{}^\epsilon_{\epsilon'}]\,=\,
(-1)^{x^*(\epsilon)+\epsilon'(x)} Q[{}^\epsilon_{\epsilon'}].
$$

The alternating tensors decompose similarly:
$$
\wedge^2(\ccH_k)\,=\,\oplus_{(\epsilon,\epsilon'),\;\epsilon'(\epsilon)=1}\;
\CC A[{}^\epsilon_{\epsilon'}],\qquad
U_{(s,x,x^*)}A[{}^\epsilon_{\epsilon'}]\,=\,
(-1)^{x^*(\epsilon)+\epsilon'(x)}A[{}^\epsilon_{\epsilon'}],
$$
where we have alternating forms (in the variables $X_\sigma,Y_\sigma$):
$$
A[{}^\epsilon_{\epsilon'}]\,:=\sum_\sigma (-1)^{\epsilon'(\sigma)} X_\sigma Y_{\sigma+\epsilon}.
$$
For example:
$$
A[{}^1_1]\,:=\,X_0Y_1\,-\,X_1Y_0,
$$
while for $k=2$:
$$
\begin{array}{l}
A[{}^{10}_{10}]\,:=\,X_{00}Y_{10}\,-\,X_{10}Y_{00}\,+
X_{01}Y_{11}\,-\,X_{11}Y_{01},\\
A[{}^{11}_{10}]\,:=\,X_{00}Y_{11}\,-\,X_{10}Y_{01}\,+
X_{01}Y_{10}\,-\,X_{11}Y_{00},\\
A[{}^{10}_{11}]\,:=\,X_{00}Y_{10}\,-\,X_{10}Y_{00}\,-
X_{01}Y_{11}\,+\,X_{11}Y_{01},\\
A[{}^{01}_{01}]\,:=\,X_{00}Y_{01}\,+\,X_{10}Y_{11}\,-
X_{01}Y_{00}\,-\,X_{11}Y_{10},\\
A[{}^{11}_{01}]\,:=\,X_{00}Y_{11}\,+\,X_{10}Y_{01}\,-
X_{01}Y_{10}\,-\,X_{11}Y_{00},\\
A[{}^{01}_{11}]\,:=\,X_{00}Y_{01}\,-\,X_{10}Y_{11}\,-
X_{01}Y_{00}\,+\,X_{11}Y_{10}.\\
\end{array}
$$
In particular, for $k=3$ the $28$ $A[{}^\epsilon_{\epsilon'}]$'s are a natural basis (of $H_3$-eigenfunctions) of $\wedge^2(\ccH_3)$. Note that the Pfaffian (a square root of the determinant of the corresponding alternating matrix) 
$pfaf(A[{}^\epsilon_{\epsilon'}])$ is non-zero for these forms.

\subsection{The Hopf maps}\label{hopf}
It was observed in \cite{LSV}, section 7, 
that there is a relation with the Hopf maps of spheres 
$$
h_k:\,S^{2^k-1}\,\longrightarrow\, S^{2^{k-1}},\qquad k=2,3,4.
$$

It is instructive to explicitly see how this comes about in the case $k=2$. One way to construct the Hopf fibration 
is to embed $S^3$ in the space of quaternions $\HH$, 
a four dimensional real vector space:
$$
S^3=\{q \in \HH :q \bar{q}=1 \}.
$$
We then identify 
$$
\RR^3=\{h \in \HH: \bar h=-h\},\qquad S^2=\{h \in \RR^3: h \bar h=1\}.
$$
With this notation a Hopf map can be defined as:
$$
h_2\,:\,S^3\,\longrightarrow\,S^2,\qquad
q\, \longmapsto\, q i \bar q
$$
which is well defined because $\bar q i q=-\overline{\bar q i q}$ is imaginary and because $\bar q q=1$ implies
$\overline{q i \bar q} q i \bar q=1$.
Recall the explicit expression for a quaternion
$q=a+b i + c j +d k$ where $i,j,k$ are the three imaginary units
(which satisfy $i^2=j^2=k^2=ijk=-1$).
Then the Hopf map is: 
$$
q=a+b i + c j +d k  \longmapsto \bar q i q= \left(a^2+b^2-c^2-d^2\right)i+ 2\left(ad+bc\right)j+2\left(bd-ac\right)k.
$$
It is easily verified that the image of $S^3$ is indeed contained in $S^2$ because:
$$
\left(a^2+b^2-c^2-d^2\right)^2+ \left(2ad+2bc\right)^2+\left(-2ac+2bd\right)^2=\left( a^2+b^2+c^2+d^2 \right)^2.
$$
On the other hand, one easily recognizes the four quadratic polynomials in $a,b,c,d$ in 
this formula as certain $Q[{}^\epsilon_{\epsilon'}]$'s, in fact the formula is equivalent
to (with $a=X_{00}, b=X_{01},c=X_{10}, d=X_{11}$):
$$
Q[{}^{00}_{10}]^2\,+\,
Q[{}^{11}_{00}]^2\,+\,Q[{}^{10}_{01}]^2\,=\,Q[{}^{00}_{00}]^2.
$$
Thus the Hopf map is given by:
$$
S^3\,\longrightarrow\,S^2,\qquad 
x\,\longmapsto\,
(Q[{}^{00}_{10}](x),\,
Q[{}^{11}_{00}](x),\, -Q[{}^{10}_{01}](x)).
$$

There exist many such quadratic relations, for example one also has:
$$
Q[{}^{00}_{00}]^2\,=\,Q[{}^{00}_{01}]^2\,+\,
Q[{}^{01}_{00}]^2\,+\,Q[{}^{11}_{11}]^2,
$$
in fact, the vector space spanned by the squares of the ten $Q[{}^\epsilon_{\epsilon'}]$'s is only 5 dimensional.
The action of the group $N_k$ from section \ref{weyl} on the set of the $Q[{}^\epsilon_{\epsilon'}]^2$, which permutes them up to sign,
allows one to find many relations from any given relation.

For $k=1$ there is an essentially unique relation:
$$
Q[{}^0_0]^2=Q[{}^0_1]^2 + Q[{}^1_0]^2
$$
The same construction extends to $k=3$ (and according to \cite{LSV} to $k=4$, which would be the case $k=3$ for them).

\

In the case $k=3$ one has the following relation:
$$
Q[{}^{000}_{000}]^2\,=\,Q[{}^{000}_{100}]^2+
Q[{}^{100}_{000}]^2+Q[{}^{101}_{101}]^2+
Q[{}^{110}_{111}]^2+Q[{}^{111}_{110}]^2.
$$
This relation does imply that there is a map
$$
S^7\longrightarrow S^4\subset \RR^5,\qquad
x\longmapsto(Q[{}^{000}_{100}](x),
Q[{}^{100}_{000}](x),Q[{}^{101}_{101}](x),
Q[{}^{110}_{111}](x),Q[{}^{111}_{110}](x))
$$
but we don't know if it is the Hopf map. Similarly, for $k=4$ we have:
$$
Q[{}^{0000}_{0000}]^2\,=\,
\sum_{i=1}^9Q[{}^{\epsilon_i}_{\epsilon_i'}]^2,\qquad
[{}^{\epsilon_i}_{\epsilon_i'}]=
[{}^{0000}_{1000}],\;
[{}^{1000}_{0000}],\;
[{}^{1001}_{1001}],\;
[{}^{1010}_{1011}],\;
[{}^{1011}_{1110}],\;
[{}^{1100}_{1111}],\;
[{}^{1101}_{1100}],\;
[{}^{1110}_{1101}],\;
[{}^{1111}_{1010}],
$$
this relation defines a map
$
S^{15}\longrightarrow S^8\;(\subset\RR^9)$.

\subsection{The Clifford algebra of $SO(2k+2)$} \label{clifford}
The two half-spin representations of $SO(2k+2)$ have dimension $2^k$
and there is in fact a close relation between these representations and the Heisenberg group. A key point is that the orthogonal group of the quadratic form $\sum_{i=1}^{2k+2} x_i^2$ has a finite subgroup $H$, isomorphic to $(\ZZ/2\ZZ)^{2k+1}$, of elements which change the signs of
an even number of the coordinates $x_1,\ldots,x_{2k+2}$. 
The inverse image $\tilde{H}$ of this finite group in the spin group $Spin(2k+2)$
is not abelian. In fact, the image of $\tilde{H}$ in the half-spin representations of the spin group is isomorphic to the Heisenberg group in its Schr\"odinger representation (cf.\ \cite{vbhe}).
Another nice feature is that the group of even permutations of the coordinates, again a subgroup of $SO(2k+2)$, will act as a subgroup of the normalizer $N_k$ of the Heisenberg group (cf.\ section \ref{weyl} for $N_k$).
See also \cite{LSV}, section 3, where a relation between the $2^k$ dimensional vector space $\ccH_k$ and a half-spin representation of $SO(2k+2)$ is considered for $k=3$ (actually, due to triality, they consider the standard representation of $SO(8)$).

\section{}
\subsection{} We consider the action of the elements $M_v$ in the normalizer of the Heisenberg group on symmetric and antisymmetric tensors. This naturally leads to the introduction of quadratic forms $q_w$ on the finite vector space
$V_k$. In the case $k=3$ some of these forms, the `odd' ones, are quite useful in understanding the root subsystems of type $E_6$ in $E_7$ and for the study of the 56-dimensional irreducible representation of $E_7$.

\subsection{The action of the transvections on tensors}\label{norten}
We determine the action of  $M_v$ on a quadric $Q[{}^\epsilon_{\epsilon'}]$ (with $\epsilon'(\epsilon)=0$) and on an alternating tensor  $A[{}^\epsilon_{\epsilon'}]$ (with $\epsilon'(\epsilon)=1$), cf.\ section \ref{tensors}. The obvious guess, that if $v=(x,x^*)\in V_k$ then $M_v(Q[{}^\epsilon_{\epsilon'}])=Q[{}^{x+\epsilon}_{x^*+\epsilon'}]$
cannot be true since one needs $(x^*+\epsilon')(x+\epsilon)=0$, else the polynomial $Q[{}^{x+\epsilon}_{x^*+\epsilon'}]$ is identically zero. 

One way to find the correct transformations is to use the explicit formula for the action of $U_v$ on  $X_\sigma$ and  
the one for $M_v$ given above. 
Instead we observe that the Heisenberg group matrices are either symmetric or antisymmetric
$$
{}^tU_v\,=\,(-1)^{x^*(x)}U_v,\qquad v=(x,x^*)\,\in V_k.
$$
Comparing this with the formula for the group law in the Heisenberg group (section \ref{Hk}) one finds:
$$
({}^tU_v)U_v\,=\,U_v({}^tU_v)\,=\,I,\qquad\mbox{so}\quad
{}^tU_v\,=\,U_v^{-1}.
$$
Moreover, using the matrices $U_v$ to define bilinear forms, we find:
$$
Q[{}^\epsilon_{\epsilon'}]\,=\,{}^tXU_{(\epsilon,\epsilon')}X,\qquad
A[{}^\epsilon_{\epsilon'}]\,=\,{}^tXU_{(\epsilon,\epsilon')}Y,\qquad
\mbox{with}\quad X:=(\ldots, X_\sigma,\ldots), Y:=(\ldots, Y_\sigma,\ldots).
$$
For example,
$$
U_{(1,1^*)}\,=\,U_{1^*}U_1=\left( \begin{array}{cc}0&1\\-1&0\end{array}\right),
\qquad
(X_0\;X_1)\left( \begin{array}{cc}0&1\\-1&0\end{array}\right)
\left(\begin{array}{l}Y_0\\Y_1\end{array}\right)\,=\,X_0Y_1-X_1Y_0\,=\,A[{}^1_1].
$$

The action of $U_v\in H_k$, with $v=(x,x^*)$, is then simply 
$$
U_{(\epsilon,\epsilon')} \longmapsto U_vU_{(\epsilon,\epsilon')}( ^tU_v)
\,=\,(-1)^{E((x,x^*),(\epsilon,\epsilon'))}U_{(\epsilon,\epsilon')}
U_v( ^tU_v)
\,=\, (-1)^{x^*(\epsilon)+\epsilon'(x)}U_{(\epsilon,\epsilon')},
$$
that is, 
$$
U_vQ[{}^\epsilon_{\epsilon'}]\,=\,
(-1)^{x^*(\epsilon)+\epsilon'(x)}Q[{}^\epsilon_{\epsilon'}],
\qquad
U_vA[{}^\epsilon_{\epsilon'}]\,=\,
(-1)^{x^*(\epsilon)+\epsilon'(x)}A[{}^\epsilon_{\epsilon'}].
$$
This agrees with the results from section \ref{tensors}.

The action of the element $M_v\in N_k$ on a tensor represented by 
$U_w$ where $w={(\epsilon,\epsilon')}$ is 
also easy to compute as $M_vU_w({}^tM_v)$.
In case $v=(x,x^*)$ with $x^*(x)=0$ we have $U_v={}^tU_v=U_v^{-1}$ and ${}^tM_v=M_v$, thus
{\renewcommand{\arraystretch}{1.7}
$$
\begin{array}{rl}
U_w\,\mapsto\,&-\frac{1}{2}(I+iU_v)U_{w}(1+iU_v)\\
=&
-\frac{i}{2}(U_{w}+i(U_vU_{w}
+U_{w}U_v)-U_vU_{w}U_v)  \\
=&
\left\{
\begin{array}{rrr}
-\frac{i}{2}(U_w+0+U_v^2U_w)=
&-iU_w &\mbox{if}\quad E(v,w)=1,\\ 
-\frac{i}{2}(U_w+2iU_vU_w-U_v^2U_w)=
&(-1)^{\epsilon'(x)}U_{v+w} 
&\mbox{if}\quad E(v,w)=0.
\end{array}
\right.
\end{array}
$$
}
Similarly, in case $v=(x,x^*)$ with $x^*(x)=1$ we find:
$$
U_w\,\mapsto\,\left\{
\begin{array}{rr}
-iU_w
&\mbox{if}\quad E(v,w)=0,\\ 
-i(-1)^{\epsilon'(x)}U_{v+w} 
&\mbox{if}\quad E(v,w)=1.
\end{array}
\right.
$$
In both cases, the results can be conveniently summarized as follows:
$$
M_v(Q[{}^\epsilon_{\epsilon'}])\,=\,
\left\{
\begin{array}{lr}
i^lQ[{}^\epsilon_{\epsilon'}]
&\mbox{if}\quad x^*(x)+E(v,w)=1,\\ 
i^lQ[{}^{\epsilon+x}_{\epsilon'+x^*}]
&\mbox{if}\quad x^*(x)+E(v,w)=0.
\end{array}
\right.
$$
for some integer $l$ which depends on $v=(x,x^*),w:=(\epsilon,\epsilon')$.
A similar result holds for the alternating forms:
$$
M_v(A[{}^\epsilon_{\epsilon'}])\,=\,
\left\{
\begin{array}{lr}
i^lA[{}^\epsilon_{\epsilon'}]
&\mbox{if}\quad x^*(x)+E(v,w)=1,\\ 
i^lA[{}^{\epsilon+x}_{\epsilon'+x^*}]
&\mbox{if}\quad x^*(x)+E(v,w)=0.
\end{array}
\right.
$$

These formulas imply that the group generated by the $M_v$ permutes the $2^{k-1}(2^k+1)$ polynomials $Q[{}^\epsilon_{\epsilon'}]^4$, in particular their sum is an invariant of degree $8$. One example already appeared in section \ref{codes}, where it was identified with the weight enumerator of the Hamming code. As the $Q[{}^\epsilon_{\epsilon'}]$ are eigenvectors for the
Heisenberg group elements $U_v$, with eigenvalues $\pm 1$, the Heisenberg group acts trivially on the $Q[{}^\epsilon_{\epsilon'}]^2$'s. Thus we get a representation of the quotient group $Sp(2k,\FF_2)$ on the vector space 
spanned by these quartic polynomials. This representation is known to be irreducible of dimension $(2^k+1)(2^{k-1}+1)/3$, cf.\ \cite{DG} and references given there.

Similarly, the $2^{k-1}(2^k-1)$ polynomials $A[{}^\epsilon_{\epsilon'}]^4$
are permuted by the group generated by the $M_v$ and this action factors over the quotient group $Sp(2k,\FF_2)$. In particular, for $k=2$ we get a homomorphism $Sp(4,\FF_2)\rightarrow S_6$, the symmetric group, which describes the permutations  of the six $A[{}^\epsilon_{\epsilon'}]^4$, see also
section \ref{n2s6}.

\subsection{Quadratic forms on $V_k$}\label{trqua}
The transformation formula for the tensors given in the previous section suggests the following definition. For $w:=(\epsilon,\epsilon')\in V_k$ define
a quadratic form $q_w$ on $V_k$ by:
$$
q_w\,:\,V_k\,\longrightarrow\,\FF_2,\qquad v=(x,x^*)\,\longmapsto\,
x^*(x)+E(v,w)=x^*(x)+x^*(\epsilon)+\epsilon'(x).
$$
Then we can rewrite the transformation formula from the preceding section as
$$
M_v(Q[{}^\epsilon_{\epsilon'}])\,=\,
\left\{
\begin{array}{ll}
i^lQ[{}^\epsilon_{\epsilon'}]
\qquad&\mbox{if}\quad q_w(v)=1,\\ 
i^lQ[{}^{\epsilon+x}_{\epsilon'+x^*}]
&\mbox{if}\quad q_w(v)=0.
\end{array}
\right.
$$

Note that $x^*(\epsilon)=\sum_i\epsilon_i x^*_i=\sum_i \epsilon_i(x^*_i)^2$
because $x^*_i=(x^*_i)^2\in \FF_2$. Thus $q_w$ is actually homogeneous of degree two(!):
$q_w(\lambda v)=\lambda^2 q_w(v)=\lambda q_w(v)$, again because $\lambda^2=\lambda\in\FF_2$.
One has, with $u=(\eta,\eta')\in V_k$:
$$
q_{u+w}(v)\,=\,x^*(x)+x^*(\epsilon+\eta)+(\epsilon'+\eta')(x)\,=\,
q_w(v)+E(v,u).
$$
One verifies by direct computation that
$$
q_w(u+v)\,=\,q_w(u)\,+\,q_w(v)\,+\,E(u,v),\qquad 
(u,v,w\in V_k),
$$
i.e., $E$ is the bilinear form associated with the quadratic form $q_w$.
A further computation shows that
{\renewcommand{\arraystretch}{1.7}
$$
\begin{array}{rcl}
q_w(t_v(u))&=&q_w(u+E(u,v)v)\\
&=& q_w(u)+q_w(E(u,v)v)+E(u,E(u,v)v)\\
&=&q_w(u)+E(u,v)q_w(v)+E(u,v)E(u,v)\\
&=&\left\{\begin{array}{lr}
q_w(u)
&\mbox{if}\quad q_w(v)=1,\\ 
q_{w+v}(u)
&\mbox{if}\quad q_w(v)=0.
\end{array}
\right. 
\end{array}
$$
}
Thus, defining an action of the involutions $t_v$ on the set of the quadrics
$q_w$, with $w\in V_k$ by
$$
t_v(q_w)\,:\,V_k\,\longrightarrow\,\FF_2,\qquad
t_v(q_w)(u)\,:=\,q_w(t_v(u))
$$
and writing $Q_{q_w}$, $A_{q_w}$ for 
$Q[{}^\epsilon_{\epsilon'}]$, $A[{}^\epsilon_{\epsilon'}]$ 
respectively, we can rewrite the transformation formula in the following way:
$$
M_v(Q_{q_w})=i^lQ_{t_v(q_w)},\qquad 
M_v(A_{q_w})=i^lA_{t_v(q_w)},
$$
with $l\in\ZZ$ depending on $v,w$.
Thus the quadratic forms $q_w$ on $V_k$ parametrize, in a natural way, the quadratic forms $ Q[{}^\epsilon_{\epsilon'}]$ and the alternating forms 
$A[{}^\epsilon_{\epsilon'}]$. This parametrization has the advantage of being compatible with the actions of $N_k$ and its quotient $Sp(2k,\FF_2)$.

\subsection{Orthogonal groups}\label{ortho}
The action of the involutions $t_v$ on the quadrics $q_w$ can be extended in an obvious way to an action of the group 
$Sp(2k,\FF_2)$ on these quadrics by defining for $g\in Sp(2k,\FF_2)$
the quadric $g(q_w)$ by $g(q_w)(v):=q_w(g^{-1}v)$.

The orthogonal group of the quadratic form $q_w$ is defined, as usual, by
$$
O(q_w)\,:=\,\{\,g\in GL(V_k)\,:\, q_w(gv)\,=\,q_w(v)\quad\forall v\in V_k\,\}.
$$
As $E$ is the associated bilinear form of $q_w$, one easily sees
that $O(q_w)\subset Sp(2k,\FF_2)$. 
See also \cite{Carter}, section 1.6 for these groups.

We showed that $t_v(q_w)=q_w$ if $q_w(v)=1$. Hence $O(q_w)$ contains the subgroup generated by the transvections $t_v$ with
$q_w(v)=1$ and in fact $O(q_w)$ is generated by these transvections.

In case $q_w(v)=0$ we showed that $t_v(q_w)=q_{v+w}$.
An amusing consequence of this result is that the action of $Sp(2k,\FF_2)$ 
on the  even (and on the odd) quadrics is transitive: given $[{}^\epsilon_{\epsilon'}]$
and $[{}^\eta_{\eta'}]$ with $\epsilon'(\epsilon)=\eta'(\eta)$,
then $v:=(\epsilon+\eta,\epsilon'+\eta')$ satisfies $q_{(\epsilon,\epsilon')}(v)=0$,
and thus $t_v(q_{(\epsilon,\epsilon')})=q_{(\eta,\eta')}$.
This argument actually shows that, for each $w=(\epsilon,{\epsilon'})\in V_k$,
there is a bijection between the sets
$$
\{v\in V_k\,:\,q_w(v)\,= 0\;\}\,=\,\{q_u\,:\;u=(\eta,\eta')\;\mbox{and}\;
\eta'(\eta)=\epsilon'(\epsilon)\,\}.
$$
In particular, an even quadratic form is zero in $2^{k-1}(2^k+1)$ points and
an odd quadratic form is zero in $2^{k-1}(2^k-1)$ points. 
The case $w=(0,0)$ is particularly obvious: if $v=(x,x^*)$ we get
$q_w(v)=x^*(x)$ and thus $q_w(v)=0$ iff $q_v$ is even.

As a consequence of the transitivity of the action of $Sp(2k,\FF_2)$ 
on the even/odd quadratic forms we get the following formulas relating the orders of the various groups, 
where $w=(\epsilon,{\epsilon'})\in V_k$:
$$
\sharp Sp(2k,\FF_2)\,=\,\left\{
\begin{array}{lr}
2^{k-1}(2^k+1)\cdot \sharp O(q_{w})
&\mbox{if}\quad \epsilon'(\epsilon)=0,\\ 
2^{k-1}(2^k-1)\cdot \sharp O(q_{w})
&\mbox{if}\quad \epsilon'(\epsilon)=1.\\ 
\end{array}
\right.
$$
More results on the action of $Sp(2g,\FF_2)$ on these quadratic forms can be found in \cite{Igusa}, Ch.\ V.6.

\

\section{}\label{C}
\subsection{} In this appendix we give some more examples of subgroups of $Sp(2k,\FF_2)$ which are defined by Coxeter relations.

\subsection{The normalizer $N_2$ and the symmetric group $S_6$}
\label{n2s6}
Using the Coxeter relations one can establish 
an explicit isomorphism between $Sp(4,\FF_2)$ and the symmetric group $S_6$. In the diagram below,  
$({}^{ab}_{cd})$ stands for  element $v=((a,b),(c,d))\in V_2$: 
$$
 
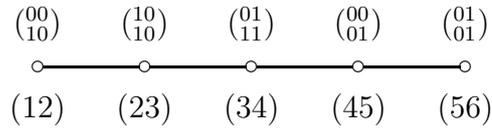
\begin{picture}(50, 95)%
   \put(-40,60){\circle{4}}%
   \put(-38, 60){\line(1, 0){36}}%
   \put(0, 60){\circle{4}}%
   \put(2, 60){\line(1, 0){36}}%
   \put(40, 60){\circle{4}}%
   \put(42, 60){\line(1, 0){36}}%
   \put(80, 60){\circle{4}}%
   \put(82, 60){\line(1, 0){36}}%
   \put(120, 60){\circle{4}}%
   \put(-40, 70){\makebox(0, 0)[b]{$({}^{00}_{10})$}}%
   \put(0, 70){\makebox(0, 0)[b]{$({}^{10}_{10})$}}%
   \put(40, 70){\makebox(0, 0)[b]{$({}^{01}_{11})$}}%
   \put(80, 70){\makebox(0, 0)[b]{$({}^{00}_{01})$}}%
   \put(120, 70){\makebox(0, 0)[b]{$({}^{01}_{01})$}}%
\put(-40, 50){\makebox(0, 0)[t]{$(12)$}}%
   \put(0, 50){\makebox(0, 0)[t]{$(23)$}}%
   \put(40, 50){\makebox(0, 0)[t]{$(34)$}}%
   \put(80, 50){\makebox(0, 0)[t]{$(45)$}}%
   \put(120, 50){\makebox(0, 0)[t]{$(56)$}}%
\end{picture}
$$
Now let $t_{ij}=t_v$ be the transvection defined by the element $v$ above the permutation $(ij)$ in the diagram.
Note that these transvections satisfy the Coxeter relations indicated by the diagram, for example if $v,w$ are connected by an edge then one checks that $E(v,w)=1$ and thus indeed $(t_vt_w)^3=I$.
The elements in the symmetric group $S_6$ written below them satisfy the same Coxeter relations.

The abstract group generated by 5 elements with these 
Coxeter relations is the Weyl group of the roots system $A_5$, that is, the symmetric group $S_6$.
As there are five $t_v$'s in $Sp(4,\FF_2)$ which satisfy the same Coxeter relations,
there is a surjective homomorphism $S_6\rightarrow H\;(\subset Sp(4,\FF_2))$ where $H$ is the subgroup generated by these five transvections; this homomorphism maps $(12)$ to $t_v$ with $v=({}^{00}_{10})$ etc., as indicated in the diagram.
In particular, this subgroup $H$ has order at most $6!=720$ and it is isomorphic to either 
$\{1\},\{\pm 1\}$ or $S_6$, but as $\sharp H\geq 5$, we must have $H\cong S_6$.
As $Sp(4,\FF_2)$ also has order $720$, it follows that $H=Sp(4,\FF_2)$, that is, these five transvections generate $Sp(4,\FF_2)$ and $S_6\cong Sp(4,\FF_2)$.
It is amusing to observe that the $2^4-1=15$ non-zero points $v\in V_2$ 
correspond to the $15$ transpositions $t_v=(ij)\in S_6$, $1\leq i<j\leq 6$.
For example, $(13)\in S_6$ can be obtained as $(13)=(12)(23)(12)$ and thus 
corresponds to $t_{12}t_{23}t_{12}=t_v$ with $v=v_{12}+v_{23}=((10),(0,0))$.

Another way to obtain this
homomorphism from $Sp(4,\FF_2)$ to $S_6$ is to use the action of elements
of $N_2$ on six labels $[{}^{\epsilon}_{\epsilon'}]$ with 
$\epsilon'(\epsilon)=1$ (cf.\ section \ref{norten}), this action factors over $Sp(4,\FF_2)$.

\subsection{Orthogonal subgroups in $Sp(6,\FF_2)$}
In the Coxeter diagram for $W(E_7)$ in section \ref{nore7}, 
in the top row, there are six $v_i$, as well as the element $v_0$ 
appearing above $\tilde{\alpha}$, the highest root,
which are all of the type $v=(x,x^*)$ with $x^*(x)=1$, that is, $q_{(0,0)}(v)=1$. 
This implies that the corresponding transvections are elements of $O(q_{(0,0)})$, cf.\ section \ref{ortho}. From the Dynkin diagram one finds that
the transvections generate a quotient of the symmetric group $S_8$ and as
$S_8$ has only three quotients, $S_8,\{\pm 1\},\{1\}$, it follows that 
$S_8\subset O(q_{(0,0)})$. One actually has equality 
(recall that $2^{k-1}(2^k+1)\cdot \sharp O(q_{(0,0)})=\sharp Sp(2k,\FF_2)$):
$$
\sharp O(q_{(0,0)})\,=\,\sharp Sp(6,\FF_2)/36\,=\,
(2^9\cdot3^4\cdot5\cdot 7)/36\,=\,8!
\qquad\mbox{hence}\quad O(q_{(0,0)})\cong S_8.
$$
Similarly, let $q_\tomega=q_{[{}^{101}_{110}]}:V_3\rightarrow \FF_2$ be the odd quadratic form
defined by
$$
q_\tomega((x,x^*))=x^*(x)+x_1+x_2+x^*_1+x^*_3,\qquad\mbox{then}\quad
q_\tomega(v_i)=1,\quad i=1,\ldots,6.
$$
Hence the subgroup of $Sp(6,\FF_2)$  generated by the corresponding transvections 
$t_{v_i}$, $1\leq i\leq 6$ is contained in $O(q_\tomega)$. 
From the diagram it is clear that 
$O(q_\tomega)$ is a quotient of $W(E_6)$. One can show that actually
(cf.\ \cite{Bb}, Exercise $\S$4.2, p229):
$$
O(q_\tomega)\,\cong\,W(E_6),
$$
and that the $t_{v_i}$, $1\leq i\leq 6$  generate $O(q_\tomega)$.

\section{} \label{D}
\subsection{} In this appendix we work out some well known results on the reduction mod two of the root lattices of $E_7$ and $E_6$. In particular we 
determine the kernel of the map $\pi:Q(E_7)\rightarrow V_3$ from section \ref{QV} and we show how root sublattices of type $Q(E_6)$ are related to odd quadratic forms on $V_3$.  We use this to rederive once more the relation
between lines in the Fano plane and the restriction of 
$V(\omega_7)$ to $SL(2,\CC)^7$.

\subsection{The root lattice $Q(E_7)$ and $V_3$}\label{appQV}
To relate the root lattice $Q(E_7)$ to the finite vector space 
$V_3\cong\FF_2^6$ we consider first of all the $\FF_2$-vector space 
$Q(E_7)/2Q(E_7)\cong \FF_2^7$. 
The bilinear scalar product $(-,-)$
descends to a well-defined bilinear map 
$$
<-,->\,:\,\Bigl(Q(E_7)/2Q(E_7)\Bigr)^2\,\longrightarrow\,\FF_2,\qquad
<\bar{\alpha},\bar{\beta}>\,:=\,(\alpha,\beta)\;\mbox{ mod}\;2,
$$
(Indeed if $\bar{\alpha}=\bar{\alpha}_1$ and $\bar{\beta}=\bar{\beta}_1$
then $\alpha=\alpha_1+2\alpha_2$, $\beta=\beta_1+2\beta_2$ for certain  $\alpha_2,\beta_2\in Q(E_7)$ and thus $(\alpha_1,\beta_1)\equiv (\alpha,\beta)
\;\mbox{mod}\;2$.) 

As $(\alpha,\alpha)\in 2\ZZ$ for any $\alpha\in Q(E_7)$
we get $<\bar{\alpha},\bar{\alpha}>=0$, so the bilinear form $<-,->$ is alternating on the odd dimensional vector space $Q(E_7)/2Q(E_7)$. Thus it must be degenerate. In fact one easily verifies 
(taking $\alpha=\alpha_i$, $i=1,\ldots,7$) that
$$
(\gamma,\alpha)\in 2\ZZ\qquad\forall \alpha\in Q(E_7),\qquad\mbox{with}\quad
\gamma:=\alpha_2+\alpha_5+\alpha_7\quad \in Q(E_7)
$$ 
but note that $\gamma\not\in 2Q(E_7)$. Thus $\bar{\gamma}\neq 0$ but $<\bar{\gamma},\bar{\alpha}>=0$ for all $\alpha\in Q(E_7)$.

Therefore, we consider the composite $\pi$ of the two quotient maps
$$
\pi\,:Q(E_7)\,\longrightarrow\,Q(E_7)/2Q(E_7)\,\longrightarrow\,
\left(Q(E_7)/2Q(E_7)\right)/\langle \bar{\gamma}\rangle,
$$
notice that 
$$
\left(Q(E_7)/2Q(E_7)\right)/\langle \bar{\gamma}\rangle\,\cong \,
Q(E_7)/\bigr(2Q(E_7)+(\gamma+2Q(E_7))\bigl).
$$

The fact that, with the notation of section \ref{nore7}, we have $(\alpha_i,\alpha_j)\equiv E(v_i,v_j)$, implies that we can describe $\pi$ explicitly as
$$
\pi\,:Q(E_7)\,\longrightarrow\, V_3\cong \FF_2^6,\qquad
\pi(\alpha_i)\,:=\,v_i.
$$
(The fact that $\pi(\gamma)=0$ corresponds to the relation $v_2+v_5+v_7=0$).
By construction, this map is compatible with scalar product on the roots 
and the  symplectic form on $\FF_2^6$:
 $$
(\alpha,\beta)\,\equiv \, E(\pi(\alpha),\,\pi(\beta))\quad\mbox{mod}\,2\qquad
(\alpha,\beta\in Q(E_7)).
$$
The images of the $63$ pairs of roots $\pm \alpha$ of $E_7$ are exactly the
$63$ non-zero elements of $V_3\cong \FF_2^6$. 

A final comment on the kernel of $\pi$:
$$
\ker(\pi)\,=\,2Q(E_7)+(\gamma+2Q(E_7)\,=\,2P(E_7),
$$
where $P(E_7)$ is the weight lattice of $E_7$, which the dual of the root lattice:
$$
P(E_7)\,:=\,
\{\,\omega\in \RR^7:\;(\omega,\alpha)\,\in\ZZ\;\forall\alpha\in Q(E_7)\,\}\,=\,
\{\omega\in \RR^7:\;(\omega,\alpha_i)\,\in\ZZ,\;i=1,\ldots,7\, \}.
$$
Indeed, as a consequence of the fact that the determinant of the Cartan matrix $((\alpha_i,\alpha_j))$ is equal to $2$, one can show that $Q(E_7)$ is a subgroup of index two in $P(E_7)$. 
As $(\gamma,\alpha)\in 2\ZZ$ for all $\alpha\in Q(E_7)$ 
we do have $\gamma/2\in P(E_7)$, but obviously $\gamma/2\not\in Q(E_7)$.
Thus we conclude that indeed $P(E_7)=Q(E_7)+(\gamma/2+Q(E_7))$
and that $Q(E_7)/2P(E_7)\cong V_3$ 
(see also \cite{Bb}, Exercise $\S$4.3, p229).

\subsection{The weights of $V(\omega_7)$ and the odd quadratic forms $q_w$}\label{wq}
We will show that there is a natural bijection between these $28$ pairs of weights in $\Pi(\omega_7)$
and the 28 quadratic forms $q_w$ on $V_3$ which are odd in the sense
that 
$$
q_w(v)=x^*(x)+E(v,w),\qquad v=(x,x^*),w=(\epsilon,\epsilon')\in V_3,\qquad \epsilon'(\epsilon)=1.
$$
(see also \cite{Bb}, Exercise $\S$4.2, p229).
In particular, these pairs of weights also correspond naturally to the 28 alternating forms $A_{q_w}=A[{}^{\epsilon}_{\epsilon'}]\in \wedge^2\ccH_3$. This again
is related to the realization
$$
V(\omega_7)\,\cong\,(\wedge^2\ccH_3)\,\oplus\,(\wedge^2\ccH_3)^*.
$$

By definition, $\omega_7$ is perpendicular to the simple roots
$\alpha_1,\ldots,\alpha_6$. The Dynkin diagram in section \ref{nore7}
shows that these roots span a root system of type $E_6$,
such a root system has $72=2\cdot 36$ roots. 
Let $\Phi\subset Q(E_7)$ be the set of these roots:
$$
\Phi\,:=\,\{\alpha\in E_7\,:\,(\omega_7,\alpha)\,=0\,\}\,\simeq \, E_6. 
$$
Under the map $\pi:Q(E_7)\rightarrow V_3$
these map to a subset of $V_3$, we will show that there is an odd quadratic form $q_\omega$ such that this subset is the set of
points
$$
\pi(\Phi)\,=\,\{v\in V_3:\;q_\omega(v)\neq 0\,\},
$$
thus we have the property (which defines $q_\omega$)
$$
(\omega_7,\alpha)=0\quad \Longleftrightarrow\quad q_\omega(\pi(\alpha))=1
\qquad(\alpha\in E_7).
$$
To see that such a $q_\omega$ exists, we observe that the root lattice of $E_6$, 
$Q(E_6)\cong \ZZ^6$, is a sublattice of root lattice of $E_7$,
$Q(E_6)\subset Q(E_7)$,
and thus under the map $\pi:Q(E_7)\rightarrow V_3$, the image of $Q(E_6)$ is all of $V_3$, $\pi(Q(E_6))=V_3$ (in fact $\gamma\not \in Q(E_6)$).
Now we put 
$$
q_\omega\,:\,V_3\,\longrightarrow\,\FF_2,\qquad 
q_\omega(\bar{\beta})\,:=\,(\beta,\beta)/2
\;\mbox{mod}\;2,\qquad\beta\in Q(E_6);
$$
this makes sense since $(\beta,\beta)=2\ZZ$ for all $\beta\in Q(E_6)$ 
and $q_\omega$ is well-defined since
$(\beta+2\beta_1,\beta+2\beta_1)=(\beta,\beta)+4((\beta,\beta_1)+(\beta_1,\beta_1))$.
Note for example that $q_\omega(\beta)=1$ for all $\beta\in E_6$ since $(\beta,\beta)=2$.

A basic property of the quadratic form $q_\omega$ is that the associated
bilinear form is $E$. Indeed, for $\beta_1,\beta_2\in Q(E_6)$ we have:
{\renewcommand{\arraystretch}{1.7}
$$
\begin{array}{rcl}
q_\omega(\bar{\beta}_1+\bar{\beta}_2)&=&
q_\omega(\overline{\beta_1+\beta_2})\\
&\equiv&
(\beta_1+\beta_2,\beta_1+\beta_2)/2\qquad\mbox{mod}\;2\\
&\equiv&
(\beta_1,\beta_1)/2+(\beta_2,\beta_2)/2+(\beta_1,\beta_2)\qquad\mbox{mod}\;2\\
&\equiv&
q_\omega(\bar{\beta}_1)+q_\omega(\bar{\beta}_2)+
E(\bar{\beta}_1,\bar{\beta}_2)\qquad\mbox{mod}\;2.
\end{array}
$$
} 
This implies that there is a $w\in V_3$ such  that 
$q_\omega(v)=x^*(x)+E(v,w)$ for all $v\in V_3$
(since $v\mapsto x^*(x)$ is also a quadratic form with associated bilinear form $E$, the map $v\mapsto x^*(x)+q_\omega(v)$ is in fact 
a linear map and hence is of the form $v\mapsto E(v,w)$ for some $w\in V_3$).

Obviously, this construction is equivariant for the action of $W(E_7)$ so if 
$s\in W(E_7)$ then $s(\omega_7)$ is a weight of the representation 
$V(\omega_7)$ and the associated odd quadratic form is $s(q_\omega)$,
which is defined using the root system of type $E_6$ in 
$s(\omega_7)^\perp\cap E_7$. 
In particular, $(\omega,\alpha)=0$ iff $q_\omega(\pi(\alpha))=1$
for all roots $\alpha$ of $E_7$.

\subsection{Example of a $q_\omega$}\label{exqo}
We determine the odd quadratic form $q_\omega$ defined by $\omega=\omega_7$. By definition of $\omega_7$ we have
$$
(\omega_7,v_i)=0\quad\mbox{for}\quad i=1,\ldots,6,\qquad
\Longrightarrow\quad q_{\omega_7}(\pi(\alpha_i))\,=\,1
\quad\mbox{for}\quad i=1,\ldots,6.
$$ 

Using the explicit description of $\pi$ (that is, $\pi(\alpha_i)=v_i$ as in the diagram in section \ref{nore7}) one finds that the unique (odd) quadratic form 
which is non-zero on $\alpha_1,\ldots,\alpha_6$ is:
$$
q_{\omega_7}=q_{[{}^{101}_{110}]}\,:\,V_3\,\longrightarrow\,\FF_2,\qquad
q_{\omega_7}((x,x^*))=x^*(x)+x_1+x_2+x^*_1+x^*_3.
$$
(Note that this odd quadratic form appeared in section \ref{nore7}.)
We already observed in section \ref{ortho}
that an odd quadratic form is non-zero in $36$ points of $V_3$, thus these are exactly the images of the roots of $E_6$:
$$
\pi(\Phi)\,=\,\{v\in V_3\,:\,q_\omega(v)\,\neq 0\,\}.
$$

\subsection{The restriction of $V(\omega_7)$ to $SL(2,\CC)^7$, bis}
\label{res727}
Recall that the map $\pi:E_7\rightarrow V_3$
maps the $126/2=63$ pairs $\pm \alpha$ of roots of $E_7$ bijectively to the $2^6-1$
non-zero points of $V_3$ in such a way that $(\alpha,\beta)=E(\pi(\alpha),\pi(\beta))$.

Let $\Pi(\omega_7)$ be the set of $56$ weights of  
the $56$-dimensional representation $V(\omega_7)$ of $E_7(\CC)$.
Under the map $\pi$ the $56/2=28$ pairs of weights $\pm \omega$ 
in $\Pi(\omega_7)$  correspond to the $28$ odd quadratic forms $q_\omega$ 
on $V_3$ in such a way that (cf.\ \ref{wq}):
$$
(\omega,\alpha)=0\quad \Longleftrightarrow\quad q_\omega(\pi(\alpha))=1
\qquad(\alpha\in E_7,\;\omega\in \Pi(\omega_7)).
$$

This information is quite useful for understanding the restriction of the $E_7(\CC)$ representation $V(\omega_7)$ to $SL(2,\CC)^7$. 
In fact, if $L$ is the Lagrangian subspace corresponding to $SL(2,\CC)^7$,
then we need to understand
the set of the $l=\pi(\alpha)\in L$ for which $(\omega,\alpha)=1$ for
a given weight $\omega\in \Pi(\omega_7)$,  
or equivalently, the set of points $l\in L$ for which $q_\omega(l)=0$. 
As the bilinear form associated to $q_\omega$ is $E$, and $E$ is zero on $L\times L$, the restriction of $q_\omega$
to $L$ is a linear form $q_\omega:L\rightarrow \FF_2$:
$$
q_\omega(l_1+l_2)\,=\,q_\omega(l_1)+q_\omega(l_2)+E(l_1,l_2)
\,=\,q_\omega(l_1)+q_\omega(l_2).
$$
As $q_\omega$ is odd, one has $q_\omega(L)=\FF_2$. This can be checked by using the $Sp(6,\FF_2)$ action to put $L$ in standard form, say 
$L=L_3\times\{0\}=\{((a,b,c),(0,0,0)\}\subset V_3$ 
and observing that if $[\epsilon,\epsilon']$ is odd 
(so $\epsilon'(\epsilon)=1$) then at least one $\epsilon'_i$ 
is non-zero and thus the restriction of $q_{[\epsilon,\epsilon']}$ 
to $L$ is non-trivial.
Thus $q_\omega$ is linear on $L$ and surjective, hence its kernel
$$
M_\omega\,:=\,
\ker(q_\omega:L\cong\FF_2^3\longrightarrow \FF_2)\quad\cong\;\FF_2^2
$$
is a linear subspace of codimension $1$ in $L$.
The three non-zero points in $M_\omega$ correspond to three copies of $sl(2)$ in $sl(2)^7$ for which $\omega$ is a weight of $V(1)$
and the remaining $7-3=4$ points of $L$ correspond to copies with the trivial representation $V(0)$.

Therefore each non-trivial summand 
$V(w_1)\boxtimes\ldots\boxtimes V(w_7)$ (see section \ref{res56})
has three of the $w_i$ equal to one and four equal to zero. The dimension of the summand is then $2^3\cdot 1^4=8$. As
$\dim V(\omega_7)=56$ it follows that there are $56/8=7$ such irreducible summands. This is also the number of linear subspaces
of dimension 2 in $L$, thus we conclude that
$$
V(\omega_7)_{|\ss}\,=\,\oplus_{M\subset L} V_M,\qquad
V_M:=V(w_1)\boxtimes\ldots\boxtimes V(w_7)\quad 
\mbox{with}\quad w_i=
\left\{\begin{array}{ll}
1\quad&\mbox{if}\;\beta_i\in M,\\
0&\mbox{if}\;\beta_i\not\in M,
\end{array}
\right.
$$
where the sum is over the two dimensional subspaces $M$ of $L$.
This is just the result of \cite{DFq}.

\end{document}